\documentclass[12pt,a4paper]{article}

\usepackage{cite}
\usepackage{latexsym,amssymb,amsmath,amscd,amsthm,amsxtra,graphicx}

\newcommand{\ba}{\begin{eqnarray}}
\newcommand{\ea}{\end{eqnarray}}
\newcommand{\bc}{\begin{center}}
\newcommand{\ec}{\end{center}}

\newcommand{\ovl}{\overline} % Long Overline
\newcommand\chiM[2]{\chi_{{}_{#1}}^{\,#2}}
\newcommand\ovlchiM[2]{\ovl\chi_{{}_{#1}}^{\,#2}}
\newcommand\tildeovlchiM[2]{\widetilde{\ovl\chi}_{{}_{#1}}^{\,#2}}
\newcommand\tildechiM[2]{\widetilde\chi_{{}_{#1}}^{\,#2}}
\newcommand\tildetildechiM[2]{\widetilde{\widetilde\chi}_{{}_{#1}}^{\,#2}}
\newcommand\transv[2]{#1_{\!{}_{\rm t}^{} #2}}
\newcommand\longit[1]{#1_{\!{}_\ell}}
\newcommand\myslash[1]{\displaystyle{\not}#1}
\newcommand{\lvac}{\langle 0 |}
\newcommand{\rvac}{| 0 \rangle}

\begin{document}

\title{Strong decays of heavy baryons in Bethe-Salpeter formalism}
\author{\\Xin-Heng Guo\thanks{xhguo@bnu.edu.cn}\,,
Ke-Wei Wei\thanks{weikw@brc.bnu.edu.cn}\,
and Xing-Hua Wu\thanks{singhwa.wu@gmail.com} \\[10pt]
{\small\it Institute of Low Energy Nuclear Physics, Beijing Normal University},
\\
{\small\it Beijing 100875, China}}
\date{\small October 8, 2007}

\maketitle

\thispagestyle{empty}
\begin{abstract}

In this paper we study the properties of diquarks (composed of $u$ and/or $d$
quarks) in the Bethe-Salpeter formalism under the covariant instantaneous
approximation. We calculate their BS wave functions and study their effective
interaction with the pion. Using the effective coupling constant among the
diquarks and the pion, in the heavy quark limit $m_Q\to\infty$, we calculate
the decay widths of $\Sigma_Q^{(*)}$ ($Q=c,b$) in the BS formalism under the
covariant instantaneous approximation and then give predictions of the decay
widths $\Gamma(\Sigma_b^{(*)}\to\Lambda_b+\pi)$.
\\


\end{abstract}

\newpage

\setcounter{page}{1}

\section{Introduction}

A light baryon composed of $u$, $d$, or $s$ quarks is a very complex three-body
system in which all the three light quarks play important roles in the dynamics
of the baryon. However, for a baryon containing a heavy quark, in the
heavy quark limit, dynamics in the baryon is greatly simplified theoretically.
In the heavy quark limit, the heavy quark effective theory (HQET) \cite{HQET}
shows that the light quarks move in an effective static color field (in the
rest frame of the heavy baryon) and can not see the spin and flavor degrees of
the heavy quark. The heavy baryon system has an extra $SU(2)_f\otimes SU(2)_s$
symmetry. Therefore, the spin and isospin of the light quarks and the heavy
quark are conserved separately.

In recent years, more and more experimental results about heavy baryons have
been reported by various experimental collaborations, e.g. the discovery and
measurements of $\Xi_{cc}^+$ \cite{pdg_2006,xi_cc_exp}, $\Sigma_b^{(*)}$
\cite{sigma_b_exp}, and $\Xi_b^-$ \cite{xi_b_exp}, etc. in addition to much more
results about baryons containing one $c$ or $b$ quark \cite{pdg_2006}. However,
the structures and properties of these baryons are not very clear, hence more
precise experimental measurements and detailed theoretical studies are urgent.

For a baryon with one heavy quark, the diquark picture has been taken in
various references, see e.g. Refs. \cite{guo-1,guo-2}. Since the spin and the
isospin of two light quarks within the heavy baryon are conserved, they can
be regarded as a two-quark system, ``diquark'', and the diquark then combines
with the heavy quark to form the heavy baryon. In this diquark picture, the
heavy baryon can be reduced to a two-body system. In this paper, we will focus
on the study of heavy baryons containing one heavy quark and two light quarks
($u$ and/or $d$). We shall not consider the effects of the isospin symmetry
violation.

\begin{figure}[hbt]
  \centering
  \includegraphics*[scale=0.6]{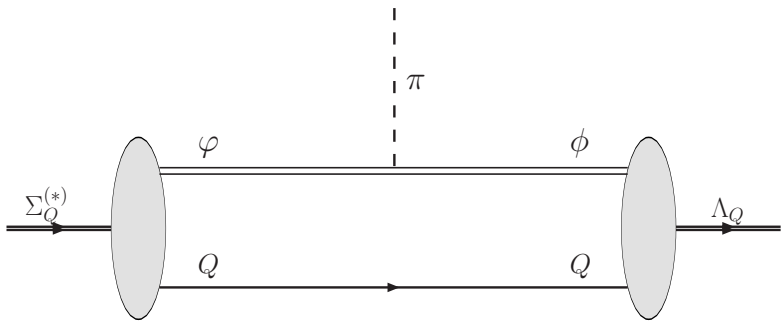}
  \caption{\label{decay-picture}
    The schematic picture for the decay processes,
    $\Sigma_Q^{(*)}\to\Lambda_Q+\pi$. The light diquark from
    $\Sigma_Q^{(*)}$ combines with the spectator heavy
    quark to form  $\Lambda_Q$ after emitting a very soft pion.
  }
\end{figure}

The purpose of this paper is to calculate the decay widths of the
strong decay processes $\Sigma_Q^{(*)}\to\Lambda_Q+\pi$ with $Q=c$
or $b$\,. These processes can be shown schematically in Fig.
\ref{decay-picture}.
%
%
%%%%%%%%%%%%%%%%%%%%%%%%%%%%%%%%%%%%%%%%%%%%%%%%%%%%%%%%%%%%%%%%%%%%%%%%%%%%%
%%%%%%%%%%%%%%%%%%%%%%%%%%%%%%%%%%%%%%%%%%%%%%%%%%%%%%%%%%%%%%%%%%%%%%%%%%%%%
%
%
From this figure, we can see that the light diquark from the
mother baryon $\Sigma_Q^{(*)}$ combines with the spectator heavy
quark to form the daughter baryon $\Lambda_Q$ after emitting a
very soft pion.

From Fig. \ref{decay-picture}, we can see that, to calculate the amplitude 
$\big\langle\Lambda_Q(P_{\!\Lambda}^{}) \pi(q)\big|\Sigma_Q^{(*)}(P_\Sigma)\big\rangle$, 
we need to know the wave functions of $\Sigma_Q^{(*)}$ and $\Lambda_Q$ and the 
effective interaction among the diquarks and the pion. In the Bethe-Salpeter
(BS) equation approach, one can show that the baryonic wave functions 
appearing in these decay amplitudes are the BS wave functions of 
$\Sigma_Q^{(*)}$ and $\Lambda_Q$ in the heavy-quark and light-diquark picture, 
which have been given in Refs. \cite{guo-1,guo-2}. On the other hand, since 
the pion emitted by the diquark is very soft, the interaction among the 
diquarks and the pion can be treated by chiral perturbation theory. In order to
calculate the (lowest order) effective coupling constant among the diquarks 
and the pion, $G_{\pi\phi\varphi}$, we will establish the BS equations for the 
scalar diquark and the axial-vector diquark, respectively. Then 
$G_{\pi\phi\varphi}$ can be expressed as the overlap integral of the BS wave 
functions of the scalar diquark and the axial-vector diquark. To simplify 
the BS equations for the diquarks to tractable forms, we will impose the 
so-called covariant instantaneous approximation in the kernels of these BS
equations. This approximation was also applied to the BS equations
for $\Sigma_Q^{(*)}$ and $\Lambda_Q$ \cite{guo-1,guo-2}. Furthermore, we will 
assume that the kernels contain a scalar confinement term and a 
one-gluon-exchange term. Throughout our calculations, we will take the heavy 
quark limit (that is to say, to neglect all the $1/m_Q$ corrections except 
for those coming from the kinematic factors).

Now let us give some more detailed explanation about the points
mentioned above. First, the decay amplitude shown in Fig. \ref{decay-picture} 
can be written out in terms of the following equation: 
\ba 
&&\big\langle\Lambda_Q(P_{\!\Lambda}^{}) \pi(q)\big|
\Sigma_Q^{(*)}(P_\Sigma)\big\rangle = \int d^4(x_1x_2y_1y_2uv)\,
\ovlchiM{P_{\!\Lambda}^{}}{}(x_2,x_1)S_Q(x_1-y_1)^{-1}
\chiM{P_\Sigma}{}{}_{,\lambda}(y_1,y_2) \nonumber\\
&&\hspace{1cm}\quad\times\,
\Delta_\phi^{-1}(x_2-u)\Delta_\varphi^{-1,\nu\lambda}(v-y_2)
\sum_{i}\langle \pi(q)|{\rm
T}\,\phi^i(u)\ovl\varphi^{\,i}_\nu(v)\rvac\,,
\label{introduction-baryon-decay-amplitude} 
\ea 
where
$P_{\!\Lambda}^{}$, $P_\Sigma$, and $q$ are the momenta of 
$\Lambda$, $\Sigma_Q^{(*)}$, and $\pi$, respectively,
$\ovlchiM{P_{\!\Lambda}^{}}{}$ and $\chiM{P_\Sigma}{}{}_{,\lambda}$ are the 
Bethe-Salpeter (BS) wave functions of $\Lambda_Q$ and $\Sigma_Q^{(*)}$, 
respectively, $\Delta_\phi$ and $\Delta_\varphi^{\nu\lambda}$ are the
propagators of the scalar diquark $\phi$ and the axial-vector diquark 
$\varphi$, respectively, $S_Q$ is the propagator of the heavy quark $Q$, 
and $i$ is the color index.

Second, let us discuss the effective coupling among the diquarks
and the pion. Since the pion emitted by the diquark is very soft,
one can expand the low energy effective Lagrangian of the diquarks
and the pion in terms of the pion's momentum. To lowest order, we
only need to consider the following interaction vertex: 
\ba 
{\cal L}_{\pi\phi\varphi}=G_{\pi\phi\varphi}^{}\sum_{b,i}\phi^i\partial^\mu\pi^b
\,\ovl\varphi_\mu^{\,b,i}+\hbox{h.c.}\,, 
\ea 
where $b$ is the isospin index. Then, the matrix element 
$\langle \pi(q)|{\rm T}\,\phi^i(u)\ovl\varphi^{\,i}_\nu(v)\rvac$ in Eq.
(\ref{introduction-baryon-decay-amplitude}) can be calculated out
from this interaction vertex, 
\ba 
\langle \pi(q)|{\rm T}\,\phi^i(u)\ovl\varphi^{\,i}_\nu(v)\rvac \sim
G_{\pi\phi\varphi}^{}\,q^\mu \int d^4 z\,e^{i q z}\,
\Delta_\phi(u-z)\Delta_\varphi(z-v)_{\mu\nu}\,. 
\ea 
On the other hand, the effective coupling constant can be related to a
transition amplitude (more details will be shown in Sect.
\ref{sect: interaction-of-qq-and-pion}), 
\ba G_{\pi\phi\varphi}^{}
\sim {1\over f_\pi}\,{\cal A}(0), 
\ea 
where 
${\cal A}(q^2)\sim q^\mu \langle P_\phi, i |A_\mu^-(q)| P_\varphi , r , i\rangle$,
which can be presented by the overlap integral of the BS wave
functions of the diquarks $\phi$ and $\varphi$. To calculate this
transition amplitude, we need the BS wave functions of $\phi$ and
$\varphi$.

The diquark systems have been studied in Ref. \cite{qq-bsEq-com} in which the
authors first solve the BS equations in the rest frame of diquarks and then
boost their solutions to a moving frame. Different from the approach of Ref.
\cite{qq-bsEq-com}, we will solve the BS equations directly under the so-called
covariant instantaneous approximation in a general coordinate system.
Furthermore, the kernel in our paper is different from that used in Ref.
\cite{qq-bsEq-com}: there are only Coulomb part (arising from
one-gluon-exchange) and the scalar linear part in our kernel,
$V\sim -{2\alpha_s\over 3r}+\kappa\, r$.

One can see from Eq. (\ref{introduction-baryon-decay-amplitude}) that,
to calculate the decay amplitudes of heavy baryons, we also need the BS
wave functions of heavy baryons. The BS wave functions of $\Lambda_Q$
and $\Sigma_Q^{(*)}$ have been studied in Refs. \cite{guo-1,guo-2}. We will
take the results given there as the input to calculate the decay amplitudes
of $\Sigma_Q^{(*)}$ in this paper.

We will constrain the ranges of parameters in our model, $m_\phi$ (or
$m_\varphi$, which is related to $m_\phi$) and $\kappa_B$, by comparing the
theoretical and experimental results about the average momentum of the heavy
quark in heavy baryons \cite{gw-07051379}. With these ranges of parameters, we
calculate the decay widths of $\Sigma_Q^{(*)}\to\Lambda_Q+\pi$.

The remainder of this paper is organized as follows. In Sect.
\ref{sect: bs-eq-of-qq}, we discuss the BS formalism under the covariant
instantaneous approximation in some details. After a brief discussion about the
BS equation of a general two-quark system, we derive the BS equations for the
scalar diquark $\phi$ and the axial-vector diquark $\varphi$, respectively.
We also give the normalization conditions of the BS wave functions in this
section. In Sect. \ref{sect: interaction-of-qq-and-pion}, we calculate the
effective coupling constant among the diquarks and the pion. In Sect.
\ref{sect-baryon-decay}, the calculation of the decay widths of
$\Sigma_Q^{(*)}\to\Lambda_Q+\pi$ are carried out in the BS formalism under the
covariant instantaneous approximation. In Sect. \ref{sect: numerical-results},
we discuss how to constrain the parameters in our model and give numerical
results for the decay widths of $\Sigma_Q^{(*)}\to\Lambda_Q+\pi$.
Sect. \ref{sect: conclusion} is reserved for our conclusions
and some discussions. We also include three appendices (\ref{sect-B},
\ref{appendix-ref-guo}, and \ref{appendix-baryon-bsw-normalization}) in this
paper. Appendix \ref{sect-B} contains some definitions. We review some results 
of Refs. \cite{guo-1,guo-2} in Appendix \ref{appendix-ref-guo}. In appendix
\ref{appendix-baryon-bsw-normalization}, we discuss the normalization
conditions of the BS wave functions of heavy baryons.

\section{Bethe-Salpeter equations for diquarks}
\label{sect: bs-eq-of-qq}

In this section, we will derive the BS equations for a scalar diquark $\phi$
and an axial-vector diquark $\varphi$. Before doing this, we will first discuss
the general BS formalism for a two-quark system (other than the system composed
of a quark and an anti-quark). The BS wave functions of the two-quark system
are defined as the following:
\ba
\chiM{P}{}(x_1,x_2)_{\alpha\beta}^{i j k}
&=& \lvac {\rm T}\,\psi_\alpha^i(x_1)\psi_\beta^j(x_2)|P,k\rangle\,,
\\
\ovlchiM{P}{}(x_2,x_1)_{\beta\alpha}^{k j i}
&=& \langle P,k |{\rm T}\,\psi_\beta^j(x_2)^*\,
\psi_\alpha^i(x_1)^* \rvac\,,
\ea
where $i,j,k$ are the color indices, $\alpha$ and $\beta$ are spinor indices,
and $P$ is the momentum of the diquark.
Note that the flavors of the quark fields are not written out explicitly, so
the flavors of the quarks could be the same or different. Since
the diquark (like an anti-quark) must furnish the representation $\ovl {\bf 3}$
of the color group $SU(3)_c$, we can define a colorless BS wave function by
$\chiM{P}{}(x_1,x_2)_{\alpha\beta}^{i j k}
={1\over 3!}\varepsilon^{i j k}\chiM{P}{}(x_1,x_2)_{\alpha\beta}$\,,
\ba
\label{diquark-bsw}
\chiM{P}{}(x_1,x_2)_{\alpha\beta}
&=& \epsilon^{ijk}\lvac {\rm T}\,\psi_\alpha^i(x_1)\psi_\beta^j(x_2)|P,k\rangle
=e^{-iPX}\int {d^4p\over (2\pi)^4} \chiM{P}{}(p)_{\alpha\beta}e^{-ipx}\,,
\\
\ovlchiM{P}{}(x_2,x_1)_{\beta\alpha}
&=& \epsilon^{ijk}\langle P,k |{\rm T}\,\psi_\beta^j(x_2)^*\,
\psi_\alpha^i(x_1)^* \rvac\,.
\label{diquark-bsw-conjugate}
\ea
where $X$ is the coordinate of the mass-center, $p$ and $x$ are the relative
momentum and the relative coordinate, respectively. As usual, we start from a
four point function,
\ba
S(x_1,x_2;y_2,y_1)^{i_1 i_2,j_2 j_1}_{\alpha_1\alpha_2,\beta_2\beta_1}
= \lvac {\rm T}\,\psi_{\alpha_1}^{i_1}(x_1)\psi_{\alpha_2}^{i_2}(x_2)
\psi_{\beta_2}^{j_2}(y_2)^*\, \psi_{\beta_1}^{j_1}(y_1)^*  \rvac \,.
%\epsilon^{i_1i_2i_3}\epsilon^{j_1j_2j_3}
\label{four-point-function}
\ea
The Fourier transform of this four point function is defined by (with indices
suppressed)
\ba
S(x_1,x_2;y_2,y_1) &=& \int {d^4P d^4P' d^4p d^4p'\over (2\pi)^{16}}
e^{-iPX+iP'Y-ipx+ip'y} \, \widetilde S(p,p',P,P')
\nonumber\\&=&
\int {d^4P d^4p d^4p'\over (2\pi)^{12}}
e^{-iP(X-Y)-ipx+ip'y} \, \widetilde S_P(p,p')\,,
\ea
where, ignoring the effects of the isospin violation,
$p=(p_1-p_2)/2,\, p'=(p_1'-p_2')/2,\, P=p_1+p_2,\, P'=p_1'+p_2'$, and
$p_1,\,p_2,\,p_1',\,p_2'$ are the momenta of
$\psi_{\alpha_1}^{i_1}(x_1)$, $\psi_{\alpha_2}^{i_2}(x_2)$,
$\psi_{\beta_1}^{j_1}(y_1)^*$, $\psi_{\beta_2}^{j_2}(y_2)^*$,
respectively. The relative coordinates and coordinates of the mass-center are
defined by $x=x_1-x_2$, $y=y_1-y_2$, $X=(x_1+x_2)/2$, and $Y=(y_1+y_2)/2$.
Then the BS wave function of a bound state composed of two quarks satisfies
the following BS equation:
\ba
\chiM{P}{}(p)^{i_1 i_2 i}_{\alpha_1\alpha_2} = \int {d^4k d^4k'\over (2\pi)^{8}}
\widetilde
S_P^{(0)}(p,k)^{i_1i_2,i_2'i_1'}_{\alpha_1\alpha_2,\alpha_2'\alpha_1'}
\widetilde K_P(k,k')^{i_1'i_2',j_2'j_1'}_{\alpha_1'\alpha_2',\beta_2'\beta_1'}
\chiM{P}{}(k')^{j_1'j_2' i}_{\beta_1'\beta_2'}\,,
\ea
where $\widetilde K_P(k,k')$ is the Fourier transform of the
two-particle-irreducible kernel of the four point function
(\ref{four-point-function})\,. In terms of the colorless BS wave function
$\chiM{P}{}(x_1,x_2)_{\alpha\beta}$\,, this equation can be written as
\ba
\chiM{P}{}(p)_{\alpha_1\alpha_2} = {1\over 6}\,\delta^{i_1i_2}_{j_1'j_2'}
\int {d^4k d^4k'\over (2\pi)^{8}}
\widetilde
S_P^{(0)}(p,k)^{i_1i_2,i_2'i_1'}_{\alpha_1\alpha_2,\alpha_2'\alpha_1'}
\widetilde K_P(k,k')^{i_1'i_2',j_2'j_1'}_{\alpha_1'\alpha_2',\beta_2'\beta_1'}
\chiM{P}{}(k')_{\beta_1'\beta_2'}\,,
\label{diquark-bsw-momentum}
\ea
where $\delta^{i_1i_2}_{j_1'j_2'} \equiv \delta^{i_1}_{j_1'}\delta^{i_2}_{j_2'}
-\delta^{i_1}_{j_2'}\delta^{i_2}_{j_1'}$\,.

To obtain the normalization conditions for the BS wave functions we also
need an inhomogeneous equation for the four point function,
\ba \label{inhomogeneous-equation}
&&\widetilde S_P(p,p')^{i_1 i_2,j_2 j_1}_{\alpha_1\alpha_2,\beta_2\beta_1}
= \widetilde S_P^{(0)}(p,p')^{i_1i_2,j_2j_1}_{\alpha_1\alpha_2,\beta_2\beta_1}
\\[5pt] &&\qquad
+ \int {d^4k d^4k'\over (2\pi)^{8}}
\widetilde
S_P^{(0)}(p,k)^{i_1i_2,i_2'i_1'}_{\alpha_1\alpha_2,\alpha_2'\alpha_1'}
\widetilde K_P(k,k')^{i_1'i_2',j_2'j_1'}_{\alpha_1'\alpha_2',\beta_2'\beta_1'}
\widetilde S_P(k',p')^{j_1'j_2',j_2j_1}_{\beta_2'\beta_1',\beta_1\beta_2} \,.
\nonumber
\ea
Near the diquark pole, we can isolate the contribution of the ``bound state''
(diquark) to the above four-point function,
\ba
\label{pole-representation-diquark}
\widetilde S_P(p,p')^{i_1i_2,j_2j_1}_{\alpha_1\alpha_2,\beta_2\beta_1}
&=& \sum_{i,j=1}^{N_c}\delta^{i j}\, {i \over P^0-E_{\bf P}-i\epsilon}\,
\chiM{P}{}(p)_{\alpha_1\alpha_2}^{i_1i_2i} \ovlchiM{P}{}(p')_{\beta_2\beta_1}^{j_2j_1j}
\nonumber\\[5pt]&&
+\, \hbox{terms regular at } P^0=E_{\bf P}\,,
\ea
where $E_{\bf P}=\sqrt{{\bf P}^2+m_{\rm diquark}^2}$ is the `on-shell'
energy of the diquark. The inverse of
$\widetilde S_P^{(0)}(p,p')$, $I_P(p,p')$, is defined to satisfy the following
equation
        \footnote{
      In general, the inversion of $S_P^{(0)}(p,p')$ will be more
      complicated. However, in our case (omitting the isospin violation),
      the following is enough.
    } :
\ba
&&\int {d^4 k\over (2\pi)^4}
\Big[ \widetilde S_P^{(0)}(p,k)\, I_P(k,p')\Big]_{\alpha_1\alpha_2, \beta_2\beta_1}^{i_1i_2,j_2j_1}
\nonumber\\[5pt]&&\quad
= {1\over 2}\,(2\pi)^4
\Big[\delta^4(p-p')\delta_{\alpha_1\beta_1}\delta_{\alpha_2\beta_2}
\delta^{i_1j_1}\delta^{i_2j_2}
\pm\delta^4(p+p')\delta_{\alpha_1\beta_2}\delta_{\alpha_2\beta_1}
\delta^{i_1j_2}\delta^{i_2j_1} \Big]\,,
\label{defintion-inverse-of-S_0}
\ea
where `+' is for the scalar diquark and `$-$' is for the axial-vector diquark.
We also need an auxiliary quantity,
\ba
&&Q_P(p,p')
\nonumber\\ &=&
\int {d^4k\over (2\pi)^4}\, (P^0-E_{\bf P}) \widetilde S_P(p,k)
{\partial\over\partial P^0}\bigg\{I_P(k,p')
-{1\over 2}\left[\widetilde K_P(k,p')\pm \widetilde K_P(-k,p')
\right]\bigg\}\,.
\nonumber
\ea
Operating this quantity upon $\chiM{P}{}$ gives the normalization condition
for the BS wave function near the diquark pole $P^0 = E_{\bf P}$ (for more
details see, e.g. Ref. \cite{lurie-book}),
\ba
&&{i\over 36}\delta^{i_1i_2}_{j_1j_2}
\int {d^4p\, d^4p'\over (2\pi)^8} \,\ovlchiM{P}{}(p)
{\partial\over\partial P^0} \bigg\{I_P(p,p')
-{1\over 2}\left[\widetilde K_P(p,p')\pm \widetilde K_P(-p,p')
\right]\bigg\}^{i_1i_2j_2j_1}\chiM{P}{}(p')
\nonumber\\[5pt]&&\quad
= 1 \,,
\ea
where the spinor indices are suppressed. Since the kernel in our
approximation is independent of $P^0$ (see the following subsections) the
normalization condition is reduced to
     \footnote{
       Eq. (\ref{diquark-normalization-condition}) is different from the
       expression in Lurie's book
       \cite{lurie-book}. The reason is that we have used a different
       convention for the normalization of one-particle state:
       $\langle {\bf p}|{\bf q}\rangle = (2\pi)^3 \delta^3({\bf p}-{\bf q})$\,.
       In deriving this normalization condition, we have used the fact that
       $\chiM{P}{}(-p)_{\beta\alpha}=-\chiM{P}{}(p)_{\alpha\beta}$
       for an iso-scalar diquark and
       $\chiM{P}{}(-p)_{\beta\alpha}=\chiM{P}{}(p)_{\alpha\beta}$
       for an iso-vector diquark which will be shown explicitly in the
       following subsections.
     },
\ba
\label{diquark-normalization-condition}
{i\over 36}\delta^{i_1i_2}_{j_1j_2}
\int {d^4p\, d^4p'\over (2\pi)^8} \,\ovlchiM{P}{}(p)
\left({\partial\over\partial P^0}I_P(p,p')^{i_1i_2j_2j_1}\right)
\chiM{P}{}(p')=1 \,,\quad
P^0 = E_{\bf P} \,.
\ea

\bigskip

Now, let us turn to the discussion about the kernel for the interaction between
quarks. The sucess of the potential model for mesons tells us that
the strong interaction between a quark and an anti-quark can be modeled
by two important interactions, a one-gluon-exchange part (with an effective
strong coupling) and a linear confinement part. We assume that this is also
true for a two-quark system. Furthermore, the parameter in the kernel of the
diquark system can be related to that of the meson system by the so-called
one-half rule \cite{half-rule}: the kernel for the diquark is just one-half
of that for the meson.

\subsection{BS equation for the scalar diquark}
\label{subsect: bse for scalar diquark}

In the following we will discuss the BS equation for a Lorentz-scalar and
isospin-scalar diquark $\phi$ composed of $d$ and $u$ quarks. The definition
of the BS wave function of the scalar diquark can be written as
\ba
\chiM{P_\phi}{}(x_1,x_2)_{\alpha\beta}
&=& \epsilon^{ijk}\,
\lvac {\rm T}\big\{d_\alpha^i(x_1)u_\beta^j(x_2)
- u_\alpha^i(x_1)d_\beta^j(x_2)\big\}|P_\phi ,k\rangle
\nonumber\\
&=&e^{-iP_\phi X}\int {d^4p\over (2\pi)^4}\, e^{-ipx}\,
\chiM{P_\phi}{}(p)_{\alpha\beta}\,.
\ea
From this definition we can see that
$\chiM{P_\phi }{}(-p)_{\beta\alpha}=-\chiM{P_\phi }{}(p)_{\alpha\beta}$\,.
The free two-particle propagators reads
\ba
\widetilde
S_{P_\phi}^{(0)}(p,p')^{i_1i_2,j_2j_1}_{\alpha_1\alpha_2,\beta_2\beta_1}
&=& 2\,(2\pi)^4\Big\{\delta^4(p-p')
[S(p_1)\gamma^0]_{\alpha_1\beta_1}^{i_1j_1}
[S(p_2)\gamma^0]_{\alpha_2\beta_2}^{i_2j_2}
\nonumber\\&&\qquad\qquad
+((\beta_1,j_1,p')\leftrightarrow(\beta_2,j_2,-p'))
\Big\}\,,
\label{forward-scattering-scalar}
\ea
where $S(p_1)$ and $S(p_2)$ are quark propagators.
The kernel arising from the one-gluon-exchange diagram is (we don't consider
the effects of the isospin violation throughout this paper)
\ba
&&\widetilde
K_{P_\phi}^{\rm (1g)}(p,p')^{i_1i_2,j_2j_1}_{\alpha_1\alpha_2,\beta_2\beta_1}
= {(ig_s)^2 \over 8}
\nonumber\\[5pt]&&\times\,
\Big\{
(\gamma^0\gamma^\mu T^a)_{\alpha_1\beta_1}^{i_1j_1}
(\gamma^0\gamma^\nu T^a)_{\alpha_2\beta_2}^{i_2j_2}
\Delta_{\mu\nu}(p'-p)
+((\beta_1,j_1,p')\leftrightarrow(\beta_2,j_2,-p'))
\Big\}\,,
\label{kernel-one-gluon-exchange-scalar}
\ea
where $T^a$ are generators of the fundamental representation of the color group
$SU(3)_c$ and $\Delta_{\mu\nu}$ is the propagator of the gluon field in Feynman
gauge. The confinement part of the kernel is assumed to be
\ba
&&
\widetilde
K_{P_\phi}^{\rm (cf)}(p,p')^{i_1i_2,j_2j_1}_{\alpha_1\alpha_2,\beta_2\beta_1}
\nonumber\\[5pt]&&\quad
= {1\over 8}\Big\{ (\gamma^0)_{\alpha_1\beta_1}^{i_1j_1}
(\gamma^0)_{\alpha_2\beta_2}^{i_2j_2} K^{\rm (cf)}(p-p')
+((\beta_1,j_1,p')\leftrightarrow(\beta_2,j_2,-p'))
\Big\}\,,
\label{kernel-confinement-scalar}
\ea
where (after imposing the covariant instantaneous approximation
\cite{meson-in-BS}, for some explanation about this approximation, see the
following text)
\ba
K^{\rm (cf)}(\transv{p}{}-\transv{p'}{})
= {c\over [-(\transv{p}{}-\transv{p'}{})^2+\mu^2]^2}
-(2\pi)^3\delta^3(\transv{p}{}-\transv{p'}{})
\int {d^3 \transv{q\,}{}\over (2\pi)^3}
{c\over [-(\transv{p}{}-\transv{q\,}{})^2+\mu^2]^2}\,,
\label{confining-potential}
\ea
where the second term is introduced to remove the infra-red
singularity of the confining kernel near the points
$\transv{p'}{} = \transv{p}{}$ and a small parameter $\mu$ is introduced to
avoid the divergence in numerical calculations. To determine the constant
$c$ in the above confinement kernel, one can compare the non-relativistic
approximation of the BS equation with the Schr\"odinger equation where the
effective potential is used in the potential model for mesons. In the
non-relativistic limit one can show that the effective potential from the BS
equation is
\ba
V(r) = -{2\over 3}{\alpha_s\over r} + {-ic\over 8\pi} r\,.
\label{cornell-potential}
\ea
Comparing Eq. (\ref{cornell-potential}) with the effective potential in the
meson case, $V(r)_{\rm meson}=-{4\over 3}{\alpha_s\over r} + \kappa r$, we can
see that setting $c=4i\pi\kappa$ is suitable from the viewpoint of the 
one-half rule \cite{half-rule}.

With the interaction kernel in Eqs. (\ref{kernel-one-gluon-exchange-scalar})
and (\ref{kernel-confinement-scalar}), the BS equation
(\ref{diquark-bsw-momentum}) becomes (using the relation
$\chiM{P_\phi}{}(-p)_{\beta\alpha}=-\chiM{P_\phi}{}(p)_{\alpha\beta}$)
\ba
\chiM{P_\phi}{}(p)&=&
\big[S(p_1)\otimes S(p_2)\big] \int {d^4p'\over(2\pi)^4}
\nonumber\\[5pt]&&\cdot\,
\Big[\gamma^\mu\otimes \gamma_\mu\, K^{\rm (1g)}(p-p')
+ 1\otimes 1 \, K^{\rm (cf)}(p-p') \Big]\cdot \chiM{P_\phi}{}(p')\,,
\ea
where $K^{\rm (1g)}={2 \over 3}g_s^2 {-i\over (p-p')^2} $\,.
For convenience, we define a deformed BS wave function,
\ba
\tildechiM{P_\phi}{}(p)_{\alpha_1\alpha_2}
=\chiM{P_\phi}{}(p)_{\gamma\alpha_2}{\cal C}^{-1}_{\gamma\alpha_1}
=({\cal C}\chiM{P_\phi}{}(p))_{\alpha_1\alpha_2}\,,
\ea
where ${\cal C}$ is the charge conjugation matrix. With this deformed BS
wave function, the BS equation becomes
\ba
\tildechiM{P_\phi}{}(p)&=&
\big[{\cal C}S(p_1){\cal C}^{-1}\otimes S(p_2)\big]
\int {d^4p'\over(2\pi)^4}
\nonumber\\[5pt]&& \cdot\,
\Big[{\cal C}\gamma^\mu{\cal C}^{-1}\otimes \gamma_\mu\, K^{\rm (1g)}(p-p')
+ 1\otimes 1 \, K^{\rm (cf)}(p-p')
\Big]\cdot \tildechiM{P_\phi}{}(p')\,.
\label{scalar-diquark-bsw-momentum}
\ea
This equation can be written in a more usual matrix form,
\ba
\tildechiM{P_\phi}{}(p)^{\rm T}=  S(p_2)  \int {d^4p'\over(2\pi)^4}
\Big[-\gamma^\mu \tildechiM{P_\phi}{}(p')^{\rm T} \gamma_\mu\, K^{\rm (1g)}
+ \tildechiM{P_\phi}{}(p')^{\rm T}\, K^{\rm (cf)} \Big] S(-p_1)\,,
\label{scalar-diquark-bsw-matrix-form}
\ea
where the superscript `T' represents the transpose of the spinor index.
In deriving the normalization condition of the BS wave function, we also need
its conjugation defined by
$\tildeovlchiM{P_\phi}{}(p)={\cal C}^{-1}\gamma^0\ovlchiM{P_\phi}{}(p)\gamma^0$,
which satisfies the following BS equation:
\ba
\tildeovlchiM{P_\phi}{}(p) = S(-p_2) \int {d^4p'\over(2\pi)^4}
\Big[-\gamma^\mu\, \tildeovlchiM{P_\phi}{}(p')\gamma_\mu\, K^{\rm (1g)}
+ \tildeovlchiM{P_\phi}{}(p')\, K^{\rm (cf)} \Big] S(p_1)\,.
\label{scalar-diquark-bsw-matrix-form-cc}
\ea
From Eqs. (\ref{scalar-diquark-bsw-matrix-form}) and
(\ref{scalar-diquark-bsw-matrix-form-cc}), we can see that
$\tildeovlchiM{-P_\phi}{}(-p)$ and $\tildechiM{P_\phi}{}(p)^{\rm T}$
satisfy completely the same equation. Therefore, once we obtain the solution for
$\tildechiM{P_\phi}{}(p)$ we also obtain the solution for its conjugate by using
$\tildeovlchiM{P_\phi}{}(p)=\tildechiM{-P_\phi}{}(-p)^{\rm T}$\,.

Now, using the decomposition of a general matrix of Dirac fields
\ba
\ovl{\psi^c}{}_{\alpha}^{\,i}\psi_{\beta}^j &=& {1\over 4}
\Big[ 1 (\ovl{\psi^c}{}^{\,i}\,\psi^j)
  +\gamma^\mu (\ovl{\psi^c}{}^{\,i}\,\gamma_\mu\psi^j )
  +{1\over 2}\sigma^{\mu\nu}(\ovl{\psi^c}{}^{\,i}\,\sigma_{\mu\nu}\psi^j)
\nonumber\\[5pt]&&\qquad
  +\,\gamma_5 (\ovl{\psi^c}{}^{\,i}\,\gamma_5\psi^j)
  -\gamma_5\gamma^\mu (\ovl{\psi^c}{}^{\,i}\,\gamma_5\gamma_\mu\psi^j )
  \Big]_{\beta\alpha}\,,
\ea
we can parametrize the BS wave functions as (note the fact that
the intrinsic parities of the quarks are the same)
\ba
\tildechiM{P_\phi}{}(p)_{\alpha_1\alpha_2}
= \Big[ \gamma_5 f_1
  + \gamma_5\gamma_\mu(P_\phi^\mu f_2 +  \transv{p}{}^{\mu} f_3)/m_\phi
 +(-i)\gamma_5\sigma_{\mu\nu}P_\phi^\mu\transv{p}{}^{\nu}f_4/m_\phi^2
  \Big]_{\alpha_2\alpha_1}\,,
\label{bsw-parametrization}
\ea
where $\longit{p}=p\cdot P_\phi/m_\phi$ is the longitudinal projection of $p$ 
along the diquark momentum 
$P_\phi$, $\transv{p}{}{}^\mu=p^\mu-\longit{p} P_\phi^\mu/m_\phi$ is 
transverse to $P_\phi$, $f_a$ $(a=1,\dots,4)$ are Lorentz-scalar functions 
of $\transv{p}{}{}^2$ and 
$\longit{p}$, $\sigma_{\mu\nu}={1\over 2}[\gamma_\mu,\gamma_\nu]$.

To simplify the BS equation (\ref{scalar-diquark-bsw-momentum}),
we impose the so-called covariant instantaneous approximation in
the kernel \cite{meson-in-BS}: $\longit{p}=\longit{p'}$\,. In this
approximation the projection of the momentum of each constituent
in the diquark along the total momentum $P_\phi$ is not changed.
For a diquark at rest, this requires the `energy' ($p_1^0$ or
$p_2^0$) of the constituent particle not to be changed. This
approximation is appropriate since the energy exchange between the
constituents in the diquark is expected to be small when we use
the constituent quark masses in the BS equation. Under this
approximation, the kernel in the BS equation is reduced to
$\widetilde K_{P_\phi}(\transv{p}{}-\transv{p'}{})$ which will be
used in the following calculations.

Furthermore, we define a three-momentum BS wave function as
$\tildechiM{P_\phi}{}(\transv{p}{})=\int{d\longit{p}\over 2\pi}\,\tildechiM{P_\phi}{}(p)$\,,
the $\longit{p'}$-integration in Eq. (\ref{scalar-diquark-bsw-momentum})
can be carried out with the following redefinitions:
\ba
\widetilde f_a(\transv{p}{})
\equiv \int {d\longit{p}\over 2\pi}\, f_a(p)\,,\quad a=1,\dots,4\,.
\ea
As usual, these four functions are not independent of each other. Their
relations can be obtained by projecting the BS wave function with
(see, e.g. Refs. \cite{lsg-rept-1991,ccw-th0312250})
\ba
\Lambda_{1,2}^\pm = {E_{1,2}\pm H_{1,2}\over 2E_{1,2}}\,,\quad
H_1 = {\myslash{{P_\phi}}\over m_\phi}(\transv{\not\! p}{}+m_1)\,,\quad
H_2 = {\myslash{{P_\phi}}\over m_\phi}(-\transv{\not\! p}{}+m_2)\,,
\ea
where the energy is defined by $E_{1,2}=\sqrt{-\transv{p}{}{}^2+m_{1,2}^2}$\,.
The square of the covariant `Hamiltonian' $H_{1,2}$ is the square of the
energy, $H_{1,2}^2=E_{1,2}^2$\,, and the projection operators satisfy
$\Lambda_f^\pm \Lambda_f^\pm = \Lambda_f^\pm$\,,
$\Lambda_f^\pm \Lambda_f^\mp = 0$\,, $f=1,2$\,.
Furthermore, we will take the quark propagators to have the form of the free
one which can be written as
\ba
\label{eq-P-slash-S-1}
{\myslash{{P_\phi}}\over m_\phi} S(p_1)
&=& i\, {m_\phi/2+\longit{p}+H_1\over
(m_\phi/2+\longit{p} - E_1 + i\epsilon)
(m_\phi/2+\longit{p} + E_1 - i\epsilon) }\,,
\\[5pt]
{\myslash{{P_\phi}}\over m_\phi} S(p_2)
&=& -i {-m_\phi/2+\longit{p}-H_2\over
(-m_\phi/2+\longit{p} + E_2 - i\epsilon)
(-m_\phi/2+\longit{p} - E_2 + i\epsilon) }\,.
\label{eq-P-slash-S-2}
\ea
Operating the projection operators on both sides of Eqs. (\ref{eq-P-slash-S-1})
and (\ref{eq-P-slash-S-2}) leads to
\begin{equation}
\Lambda_1^\pm {\myslash{{P_\phi}}\over m_\phi} S(p_1)
=i\left\{
\begin{array}{l}
\displaystyle{1\over m_\phi/2+\longit{p}-E_1 + i\epsilon}\Lambda_1^+
\\[10pt]
\displaystyle{1\over m_\phi/2+\longit{p}+E_1 - i\epsilon}\Lambda_1^-
\end{array} \right.   \,,
\end{equation}
and
\begin{equation}
\Lambda_2^\pm {\myslash{{P_\phi}}\over m_\phi} S(p_2)
=-i\left\{
\begin{array}{l}
\displaystyle{1\over -m_\phi/2+\longit{p} + E_2 - i\epsilon}\Lambda_2^+
\\[10pt]
\displaystyle{1\over -m_\phi/2+\longit{p} - E_2 + i\epsilon}\Lambda_2^-
\end{array}\right.   \,.
\end{equation}

Now, by multiplying $\Lambda_1^\pm\myslash{{P_\phi}}$ and
$\Lambda_2^\mp\myslash{{P_\phi}}$ on both sides of equation
(\ref{scalar-diquark-bsw-momentum}) and integrating out the longitudinal
momentum $\longit{p}$ along proper contour(s), we can obtain the following
constraint equations:
\ba
\label{scalar-diquark-constraint-condition-matrix-1}
(\Lambda_1^+ \myslash{{P_\phi}})_{\alpha_1'\alpha_1}
(\Lambda_2^- \myslash{{P_\phi}})_{\alpha_2'\alpha_2}
\tildechiM{P_\phi}{}(\transv{p}{})_{\gamma\alpha_2} {\cal C}_{\gamma\alpha_1}
= 0\,,
\\
(\Lambda_1^- \myslash{{P_\phi}})_{\alpha_1'\alpha_1}
(\Lambda_2^+ \myslash{{P_\phi}})_{\alpha_2'\alpha_2}
\tildechiM{P_\phi}{}(\transv{p}{})_{\gamma\alpha_2} {\cal C}_{\gamma\alpha_1}
= 0\,.~
\label{scalar-diquark-constraint-condition-matrix-2}
\ea
Substituting the parametrization Eq. (\ref{bsw-parametrization}) into
Eqs. (\ref{scalar-diquark-constraint-condition-matrix-1}) and
(\ref{scalar-diquark-constraint-condition-matrix-2}) we obtain the following
constraint relations
    \footnote{
      The above two equations can be written in matrix form:
      $\Lambda_2^-\myslash{{P_\phi}}\tildechiM{P_\phi}{}(p)^T
      {\cal C}(\Lambda_1^+\myslash{{P_\phi}})^T{\cal C}^{-1}=0$ and
      $\Lambda_2^+\myslash{{P_\phi}}\tildechiM{P_\phi}{}(p)^T
      {\cal C}(\Lambda_1^-\myslash{{P_\phi}})^T{\cal C}^{-1}=0$\,.
      These two equations are in fact linear combinations of independent
      matrices: 1, $\myslash{{P_\phi}}$, $\transv{\not\! p}{}$, and
      $\myslash{{P_\phi}}\!\transv{\not\! p}{}$\,. From the equations of
      the coefficients of these matrices we can obtain the following
      consistent solutions.
    }:
\ba
\widetilde f_3 = 0\,,\qquad \widetilde f_4 = -{m_\phi \over m}
\widetilde f_2\,,%\qquad \widetilde f_1\hbox{\quad not constrained}
\label{scalar-diquark-constraint-condition}
\ea
where we have defined $m\equiv m_1=m_2$ and consequently
$\omega\equiv E_1=E_2$\,.

In addition to the above constraint relations, we can also obtain other two
equations by operating $\Lambda_1^\pm\myslash{{P_\phi}}$ and
$\Lambda_2^\pm\myslash{{P_\phi}}$ upon both sides of equation
(\ref{scalar-diquark-bsw-momentum}):
\ba
&&\left[\Lambda_1^\pm{\myslash{P_\phi}\over m_\phi}\otimes
\Lambda_2^\pm{\myslash{P_\phi}\over m_\phi}\right]\cdot
\chiM{P_\phi}{}(p) =
\left[\Lambda_1^\pm{\myslash{{P_\phi}}\over m_\phi}S(p_1)\otimes
\Lambda_2^\pm{\myslash{{P_\phi}}\over m_\phi} S(p_2)\right]
\nonumber\\&&\qquad
\cdot\, \int {d^4p'\over(2\pi)^4}
\Big[\gamma^\mu\otimes \gamma_\mu\, K^{\rm (1g)}(p-p')
+ 1\otimes 1 \, K^{\rm (cf)}(p-p')
\Big]\cdot \chiM{P_\phi}{}(p')\,,
\ea
which, after taking the covariant instantaneous approximation and
completing the integration $\int{d \longit{p}\over 2\pi}$ on both sides,
leads to (using $\chiM{}{}={\cal C}^{-1}\tildechiM{}{}$ and written in terms of
the matrix form)
\ba
\label{scalar-BSW-four-momentum-form}
&&-\,\Lambda_2^\pm{\myslash{{P_\phi}}\over m_\phi}
\tildechiM{P_\phi}{}(\transv{p}{})^T
{\myslash{{P_\phi}}\over m_\phi}{\cal C}(\Lambda_1^\pm)^T
= {i \over \pm m_\phi - 2 \omega}
\\&& \times\,
\Lambda_2^\pm\int {d^3\transv{p'}{}\over(2\pi)^3}
\left[-\gamma^\mu\tildechiM{P_\phi}{}(\transv{p'}{})^T \gamma_\mu
\, K^{\rm (1g)}(\transv{p}{}-\transv{p'}{})
+\tildechiM{P_\phi}{}(\transv{p'}{})^T K^{\rm (cf)}(\transv{p}{}-\transv{p'}{})
\right]
{\cal C}(\Lambda_1^\pm)^T\,.
\nonumber
\ea
Multiplying both sides of Eq. (\ref{scalar-BSW-four-momentum-form}) by
${\cal C}^{-1}\gamma_5$ from the right and taking the trace over
the spinor indices gives
\ba
\label{scalar-component-bsw-3d-1}
\widetilde f_1(\transv{p}{}) &=& {1\over \omega(4\omega^2-m_\phi^2)}
\int{d^3\transv{p'}{}\over(2\pi)^3}\,\bigg\{
2\omega^2(V^{(\rm cf)}+4 V^{(\rm 1g)})\widetilde f_1(\transv{p'}{})
\nonumber\\&&\qquad\qquad\qquad
-\,{m_\phi\over m}\Big[(m^2+\transv{p}{}\cdot\transv{p'}{})V^{(\rm cf)}
-2m^2V^{(\rm 1g)}\Big]\widetilde f_2(\transv{p'}{})
\bigg\}\,,
\\[5pt]
\widetilde f_2(\transv{p}{}) &=& {1\over \omega(4\omega^2-m_\phi^2)}
\int{d^3\transv{p'}{}\over(2\pi)^3}\,\bigg\{
-mm_\phi (V^{(\rm cf)}+4 V^{(\rm 1g)})\widetilde f_1(\transv{p'}{})
\nonumber\\&&\qquad\qquad\qquad
+\,2\Big[(m^2+\transv{p}{}\cdot\transv{p'}{})V^{(\rm cf)}
-2m^2V^{(\rm 1g)}\Big]\widetilde f_2(\transv{p'}{})
\bigg\}\,.
\label{scalar-component-bsw-3d-2}
\ea
where $V^{(\rm cf)}=-iK^{(\rm cf)}$ and $V^{(\rm 1g)}=-iK^{(\rm 1g)}$.
At this point, let us define three-vectors $\transv{{\bf p}'}{}$ and
$\transv{{\bf p}}{}$ which satisfy
$\transv{p'}{}{}^2=-\transv{{\bf p}'}{}{}^2$\,,
$\transv{p}{}^2=-\transv{{\bf p}}{}^2$ and
$\transv{p'}{}\cdot\transv{p}{}=-\transv{{\bf p}}{}\cdot\transv{{\bf p}'}{}$
     \footnote{
       Since $\transv{p'}{}$ and $\transv{p}{}$ are four-vectors perpendicular
       to the total momentum ${P_\phi}$\,, which is a time-like four-vector, we
       can always accomplish this.
     }.
Using these definitions, after completing the azimuthal integration, we can
rewrite Eqs. (\ref{scalar-component-bsw-3d-1}) and
(\ref{scalar-component-bsw-3d-2}) in the one-dimensional integration form,
\ba
\widetilde f_a(|\transv{{\bf p}}{}|)={1\over \omega(4\omega^2-m_\phi^2)}
\int d |\transv{{\bf p}'}{}| \sum_{b=1,2}\Big[
A_{ab}(|\transv{{\bf p}}{}|, |\transv{{\bf p}'}{}|)
\widetilde f_b(|\transv{{\bf p}'}{}|)
-D_{ab}(|\transv{{\bf p}}{}|, |\transv{{\bf p}'}{}|)
\widetilde f_b(|\transv{{\bf p}}{}|)   \Big]\,,
\ea
where $a,b=1,2$\,, the `matrices' are defined by
\ba
A_{11} &=& 2\omega^2(L_0+4 G_0)\,,\qquad
A_{12} = - {m_\phi \over m}(m^2 L_0+L_1-2m^2G_0 )\,,
\nonumber\\
A_{21} &=& -\, mm_\phi( L_0+4G_0)\,,\qquad
A_{22} = 2( m^2L_0+L_1-2m^2G_0 )\,,
\nonumber
\ea
and the counter terms are
\ba
D_{11} &=& 2\omega^2 L_0\,,\qquad
D_{12} = - {m_\phi \over m}(m^2-|\transv{{\bf p}}{}|^2) L_0\,,
\nonumber\\
D_{21} &=& -\, mm_\phi L_0 \,,\qquad
D_{22} = 2 (m^2-|\transv{{\bf p}}{}|^2) L_0\,,
\nonumber
\ea
where the definitions of $L_0$\,, $L_1$, and $G_0$ are given in Appendix
\ref{sect-B}\,. The parameter $\mu$ will be removed in the end of
the calculation by letting $\mu\to 0$ (in fact, taking $\mu$ to be sufficiently
small is enough for pratical calculations).

\subsection{Normalization condition for the BS wave function of the scalar
diquark}

Now, we will discuss the normalization condition for the BS wave function
of the scalar diquark $\phi$. From Eqs. (\ref{defintion-inverse-of-S_0}) and
(\ref{forward-scattering-scalar}) we
can define the inverse of $\widetilde S^{(0)}$ as
\ba
I_P(p,p')^{i_1i_2,j_2j_1}_{\alpha_1\alpha_2,\beta_2\beta_1}
= {1\over 4}\delta^{i_1j_1}\delta^{i_2j_2}(2\pi)^4\delta^4(p-p')
[S(p_1)\gamma^0]^{-1}_{\alpha_1\beta_1}
[S(p_2)\gamma^0]^{-1}_{\alpha_2\beta_2}\,.
\ea
From Eq. (\ref{diquark-normalization-condition}) we have the normalization
condition for the BS wave function,
\ba
{i\over 24}\int {d^4p\over (2\pi)^4} \,\ovlchiM{P_\phi}{}(p)_{\beta_2\beta_1}
{\partial\over\partial P^0}  \Big\{[S(p_1)\gamma^0]^{-1}_{\alpha_1\beta_1}
[S(p_2)\gamma^0]^{-1}_{\alpha_2\beta_2}\Big\}
\chiM{P_\phi}{}(p)_{\alpha_1\alpha_2} = 1\,.
\label{scalar-norm-44-momentum}
\ea
This equation can be recast in the matrix form,
\ba
&&{1\over 48}\int {d^4p\over(2\pi)^4}\,{\rm Tr}\,
\bigg\{{\cal C}\gamma^0 \,\ovlchiM{P_\phi}{}(p)^{\rm T}\,\gamma^0
{\not\!\xi}\, \tildechiM{P_\phi}{}(p)^{\rm T} S(-p_1)^{-1}
\nonumber\\&&\qquad\qquad\qquad
-\, {\cal C}\gamma^0\, \ovlchiM{P_\phi}{}(p)^{\rm T}\, \gamma^0
S(p_2)^{-1}\tildechiM{P_\phi}{}(p)^{\rm T} {\not\!\xi}\, \bigg\}
= 1\,,
\ea
where $\xi=(1,\overrightarrow{0})$\,. Furthermore, we can separate out the
longitudinal momentum by using Eq. (\ref{scalar-diquark-bsw-matrix-form}):
$\tildechiM{P_\phi}{}(p)^{\rm T}
=i S(p_2)\tildetildechiM{P_\phi}{}(\transv{p}{}) S(-p_1)$ and define
\ba
\tildetildechiM{P_\phi}{}(\transv{p}{})\equiv \gamma_5\bigg\{
h_1(\transv{p}{})+{{\not\!\!P_\phi}\over m_\phi}
h_2(\transv{p}{})
- i\sigma_{\mu\nu}{P_\phi^\mu \transv{p}{}^\nu\over m_\phi^2}
h_4(\transv{p}{})\bigg\}\,,
\ea
where
\ba
h_1(\transv{p}{}) &=&
\int {d^3\transv{p'}{}\over(2\pi)^3}\,
\Big[4 V^{(\rm 1g)}(\transv{p}{}-\transv{p'}{})
+V^{(\rm cf)}(\transv{p}{}-\transv{p'}{}) \Big]
{\widetilde f}_1(\transv{p'}{})\,,
\\
h_2(\transv{p}{}) &=&
\int {d^3\transv{p'}{}\over(2\pi)^3}\,
\Big[-2 V^{(\rm 1g)}(\transv{p}{}-\transv{p'}{})
+V^{(\rm cf)}(\transv{p}{}-\transv{p'}{}) \Big]
{\widetilde f}_2(\transv{p'}{})\,,
\\
h_4(\transv{p}{}) &=&
\int {d^3\transv{p'}{}\over(2\pi)^3}\,
V^{(\rm cf)}(\transv{p}{}-\transv{p'}{})
{\transv{p}{}\cdot\transv{p'}{}\over\transv{p}{}{}^2}
{\widetilde f}_4(\transv{p'}{})\,.
\ea
On the other hand, from Eq. (\ref{scalar-diquark-bsw-matrix-form-cc})
and the discussion following that equation, we have
\ba
{\cal C}\gamma^0 \,\ovlchiM{P_\phi}{}(p)^{\rm T}\,\gamma^0
= {\cal C}\big[\tildechiM{-P_\phi}{}(-p)^{\rm T}\big]^{\rm T}{\cal C}^{-1}
\equiv i S(-p_1)\tildetildechiM{P_\phi}{}(\transv{p}{})^{(\rm c)} S(p_2)\,,
\label{scalar-diquark-factorized-cc}
\ea
where
\ba
\tildetildechiM{P_\phi}{}(\transv{p}{})^{(\rm c)} = \bigg\{
h_1(\transv{p}{})+{{\not\!\!P_\phi}\over m_\phi}
h_2(\transv{p}{})
+ i\sigma_{\mu\nu}{P_\phi^\mu \transv{p}{}^\nu\over m_\phi^2}
h_4(\transv{p}{})\bigg\}\gamma_5\,.
\ea
With these definitions, we can write the normalization condition as
\ba
&&-\,{1\over 48}\int {d^4p\over(2\pi)^4}\,{\rm Tr}\,
\bigg\{S(-p_1)\,\tildetildechiM{P_\phi}{}(\transv{p}{})^{(\rm c)}\,
S(p_2){\not\!\xi}\,S(p_2)\tildetildechiM{P_\phi}{}(\transv{p}{})
\nonumber\\&&\qquad\qquad\qquad
-\, S(-p_1)\,\tildetildechiM{P_\phi}{}(\transv{p}{})^{(\rm c)}\,
S(p_2)\,\tildetildechiM{P_\phi}{}(\transv{p}{})\,S(-p_1){\not\!\xi} \bigg\}
= 1\,.
\ea
After integrating out the longitudinal momentum $\longit{p}$ and carrying out
the trace calculation, we have the following
one-dimensional integration equation:
\ba
&&{2E_\phi\over 6m_\phi^2} \int {|\transv{{\bf p}}{}|^2d |\transv{{\bf p}}{}|
\over 2\pi^2\omega(m_\phi^2-4\omega^2)^2}
\bigg\{  2 |\transv{{\bf p}}{}|^4 (h_4)^2 - |\transv{{\bf p}}{}|^2(m_\phi^2
+ 4\omega^2)h_1 h_4 + 2m^2m_\phi^2 (h_2)^2
\nonumber\\&&\qquad
+\, 2m_\phi^2\omega^2 (h_1)^2 + 4mm_\phi |\transv{{\bf p}}{}|^2 h_2 h_4
- mm_\phi(m_\phi^2+4\omega^2)h_1 h_2
\bigg\} = 1\,,
\label{scalar-diquark-normalization-eq}
\ea
where $E_\phi=P_\phi\cdot\xi=P_\phi^0$ is the energy of the scalar diquark.

\subsection{BS equation for the axial-vector diquark}
\label{subsect: bse for axial-vector diquark}

Now we will derive the BS equation for the axial-vector diquark. Since we 
will not take the isospin violation into
account, we can take the diquark composed of $u u$ with $I_3=+1$ as example
in the following discussion. The BS wave function is defined by
\ba
\label{axial-diquark-bsw}
\chiM{P_\varphi}{(r)}(x_1,x_2)_{\alpha\beta}
&\makebox[0pt]{=}&
\epsilon^{ijk}\lvac {\rm T}\,u_\alpha^i(x_1)u_\beta^j(x_2)|P_\varphi,r,k\rangle
=e^{-i P_\varphi X}
\int{d^4p\over (2\pi)^4} \chiM{P_\varphi}{}(p)_{\alpha\beta}\,e^{-ipx}
\,,\quad
\\
\ovlchiM{P_\varphi}{(r)}(x_2,x_1)_{\beta\alpha}
&\makebox[0pt]{=}& \epsilon^{ijk}\langle P_\varphi,r,k |
{\rm T}\,u_\beta^j(x_2)^*\,u_\alpha^i(x_1)^* \rvac\,,
\label{axial-diquark-bsw-conjugate}
\ea
where $i,j,k$ are color indices, $r$ is the index of the polarization vector 
of the axial-vector diquark. The free two-particle propagator reads
\ba
\widetilde
S_{P_\varphi}^{(0)}(p,p')^{i_1i_2,j_2j_1}_{\alpha_1\alpha_2,\beta_2\beta_1}
&=& (2\pi)^4\Big\{\delta^4(p-p')
[S(p_1)\gamma^0]_{\alpha_1\beta_1}^{i_1j_1}
[S(p_2)\gamma^0]_{\alpha_2\beta_2}^{j_2j_2}
\nonumber\\&&\qquad\qquad
-(\beta_1\leftrightarrow\beta_2, j_1\leftrightarrow j_2, p'\leftrightarrow -p')
\Big\}\,.
\label{forward-4-point-axial-vector-diquark}
\ea
The kernel arising from the one-gluon exchange diagram is
\ba
&&\widetilde
K_{P_\varphi}^{\rm (1g)}(p,p')^{i_1i_2,j_2j_1}_{\alpha_1\alpha_2,\beta_2\beta_1}
= {(ig_s)^2\over 4}
\nonumber\\[5pt]&&\quad\times\,
\Big\{
(\gamma^0\gamma^\mu)_{\alpha_1\beta_1}^{i_1i_2}
(\gamma^0\gamma^\nu)_{\alpha_2\beta_2}^{j_1j_2}
\Delta_{\mu\nu}(p'-p)
-(\beta_1\leftrightarrow\beta_2, j_1\leftrightarrow j_2, p'\leftrightarrow -p')
\Big\}\,.
\ea
On the other hand, the confinement kernel is assumed to be
\ba
&&\widetilde
K_{P_\varphi}^{\rm (cf)}(p,p')^{i_1i_2,j_2j_1}_{\alpha_1\alpha_2,\beta_2\beta_1}
\nonumber\\[5pt] &&\quad
= \,{i\over 4}\,\Big\{
(\gamma^0)_{\alpha_1\beta_1}^{i_1i_2}
(\gamma^0)_{\alpha_2\beta_2}^{j_1j_2}V^{\rm (cf)}(p-p')
-(\beta_1\leftrightarrow\beta_2, j_1\leftrightarrow j_2, p'\leftrightarrow -p')
\Big\}\,.
\label{kernel-confinement-axial-vector}
\ea
From the above kernel and $\widetilde S^{(0)}$ we can obtain the BS equation
for the axial-vector diquark,
\ba
\tildechiM{P_\varphi}{(r)}(p)
&=& \big[{\cal C}S(p_1){\cal C}^{-1}\otimes S(p_2)\big]
\int {d^4p'\over(2\pi)^4}
\nonumber\\[5pt]&&\cdot\,
\Big[{\cal C}\gamma^\mu{\cal C}^{-1}\otimes \gamma_\mu\, K^{\rm (1g)}(p-p')
+ 1\otimes 1 \, K^{\rm (cf)}(p-p') \Big]\cdot
\tildechiM{P_\varphi}{(r)}(p')\,,
\label{axial-vector-diquark-bsw-momentum}
\ea
which has the same form as that for the scalar diquark, Eq.
(\ref{scalar-diquark-bsw-momentum}). Similar to the case of the scalar diquark,
we can parametrize the BS wave function of the axial-vector diquark
by several components $g_f^{}$ ($f=1,\dots, 8$), which are functions
of $\transv{p}{}$ and $P_\varphi$:
%
%      \footnote{
%   Note the fact that the state of the axial-vector diquark
%        $|P,r,k\rangle$ can be written as
%        $\varepsilon^{(r)}_\mu \varphi^{k,\mu}(P)|0\rangle$\,,
%        where $\varphi^{k,\mu}$ is the field operator of the axial-vector
%        diquark.
%      }:
%
\ba
\tildechiM{P_\varphi}{(r)}(p)_{\alpha_2\alpha_1}
&=&\varepsilon^{(r)}_\rho\,
\bigg\{ \transv{p}{}^\rho \, {g_1^{}\over m_\varphi}
+ \Big[ \transv{p}{}^\rho P^\mu {g_2^{}\over m_\varphi^2}
+ \transv{p}{}^\rho \transv{p}{}^\mu {g_3^{}\over m_\varphi^2}
\Big]\gamma_\mu
+ \transv{p}{\mu} P_\nu \gamma^{\rho\mu\nu} {g_5^{}\over m_\varphi^2}
+ \gamma^\rho g_7^{}
\nonumber\\[5pt]&&\qquad
-\,i \Big[\transv{p}{}^\rho \transv{p}{}^\mu P^\nu {g_4^{}\over m_\varphi^3}
+ g^{\rho\mu}P^\nu {g_6^{}\over m_\varphi}
+ g^{\rho\mu}\transv{p}{}^\nu {g_8^{}\over m_\varphi}
\Big]\sigma_{\mu\nu} \bigg\}_{\alpha_1\alpha_2}\,,
\label{axial-diquark-bsw-expansion}
\ea
where $\varepsilon^{(r)}_\rho$ is the $r$-th polarization state of the
axial-vector diquark, which satisfies
$\varepsilon^{(r)}_\rho P_\varphi^\rho=0$\,.
%and $|\transv{p}{}|=\sqrt{-\transv{p}{}^2}$ is the norm of $\transv{p}{}$\,.
Performing the same procedure as that for the scalar diquark in the previous
subsection, we have the following constraint relations for the coefficient
functions:
\ba
\widetilde g_2^{} = \widetilde g_8^{} \equiv 0\,,\qquad
\widetilde g_1^{} =-\,{\transv{p}{}^2\,\widetilde g_3^{}
+m_\varphi^2\,\widetilde g_7^{}\over m_\varphi m}\,,\qquad
\widetilde g_6^{} = {m\over m_\varphi }\, \widetilde g_5^{} \,.
\ea
With these constraint relations, the BS equation for the axial-vector diquark
is composed of the following four integral equations:
\ba
\widetilde g_7 &=&
{1\over 2\omega m_\varphi^2\transv{p}{}{}^2(m_\varphi^2-4\omega^2)}
\int {d^3\transv{p'}{}\over (2\pi)^3}\Bigg\{
mm_\varphi\Big[\transv{p}{}{}^2\transv{p'}{}{}^2
-(\transv{p}{}\cdot\transv{p'}{})^2\Big]V^{(\rm cf)} \widetilde g_4
\nonumber\\[5pt] &&\quad
-\, 4m_\varphi^2\transv{p}{}^2\omega^2 (V^{(\rm cf)} +2V^{(\rm 1g)})\widetilde g_7
%\nonumber\\[5pt] &&\quad
+ 2m_\varphi^2\transv{p}{}^2\Big[(m^2+\transv{p}{}\cdot\transv{p'}{}) V^{(\rm cf)}
-2(\transv{p}{}\cdot\transv{p'}{}) V^{(\rm 1g)}\Big]\widetilde g_5
\nonumber\\[5pt] &&\quad
-\, 2\omega^2\Big[\transv{p}{}^2 \transv{p'}{}{}^2
- (\transv{p}{}\cdot\transv{p'}{})^2\Big]
(V^{(\rm cf)}+2V^{(\rm 1g)}) \widetilde g_3 \Bigg\} \,,
\label{axial-bsw-component-1}
\ea
%\\[5pt]  %%%%%%%%%%%%%%%%%%%%%%%%%%%%%%%%%%%%%%%%%%%%%%%%%%%%%%%%%%%%%
\ba
\widetilde g_3 &=& {-1\over 2\omega\transv{p}{}{}^4(m_\varphi^2-4\omega^2)}
\int {d^3\transv{p'}{}\over (2\pi)^3}\Bigg\{
mm_\varphi(\transv{p}{}{}^2\transv{p'}{}{}^2
-3(\transv{p}{}\cdot\transv{p'}{})^2)V^{(\rm cf)} \widetilde g_4
\nonumber\\[5pt] &&\quad
+\, 4m_\varphi^2\transv{p}{}^2\Big[(\transv{p}{}^2
+(\transv{p}{}\cdot\transv{p'}{}))V^{(\rm cf)}
+2(\transv{p}{}^2-2(\transv{p}{}\cdot\transv{p'}{}))V^{(\rm 1g)}\Big]
\widetilde g_7
\nonumber\\[5pt] &&\quad
+\, 2m_\varphi^2 \transv{p}{}^2 (\transv{p}{}\cdot\transv{p'}{})
(V^{(\rm cf)}-2 V^{(\rm 1g)}) \widetilde g_5
\nonumber\\[5pt] &&\quad
+\, \bigg[\Big(-2 \omega^2\transv{p}{}^2 \transv{p'}{}{}^2
+4 \transv{p}{}^2 \transv{p'}{}{}^2 (\transv{p}{}\cdot\transv{p'}{})
+2(2m^2+\omega^2)(\transv{p}{}\cdot\transv{p'}{})^2
\Big)V^{(\rm cf)}
\nonumber\\ &&\quad
+\, 4\Big(-\omega^2\transv{p}{}^2 \transv{p'}{}{}^2
-4\transv{p}{}^2\transv{p'}{}{}^2(\transv{p}{}\cdot\transv{p'}{})
+(2m^2+\omega^2)(\transv{p}{}\cdot\transv{p'}{})^2
 \Big)V^{(\rm 1g)}\bigg]\widetilde g_3 \Bigg\} \,,
\label{axial-bsw-component-2}
\ea
%\\[5pt]  %%%%%%%%%%%%%%%%%%%%%%%%%%%%%%%%%%%%%%%%%%%%%%%%%%%%%%%%%%%%%
\ba
\widetilde g_4 &=& {1\over 2m\omega\transv{p}{}{}^4(m_\varphi^2-4\omega^2)}
\int {d^3\transv{p'}{}\over (2\pi)^3}\Bigg\{
2m\Big[m^2\transv{p}{}{}^2\transv{p'}{}{}^2
-(m^2+2\omega^2)(\transv{p}{}\cdot\transv{p'}{})^2\Big]V^{(\rm cf)} \widetilde g_4
\nonumber\\[5pt] &&\quad
+\, 4m^2m_\varphi\transv{p}{}^2\Big[(\transv{p}{}^2
+(\transv{p}{}\cdot\transv{p'}{}))V^{(\rm cf)}
-2(\transv{p}{}\cdot\transv{p'}{})V^{(\rm 1g)}\Big]\widetilde g_5
\nonumber\\[5pt] &&\quad
+\, 2m_\varphi^3\transv{p}{}^2(\transv{p}{}\cdot\transv{p'}{})
(V^{(\rm cf)}-4V^{(\rm 1g)})\widetilde g_7
\nonumber\\[5pt] &&\quad
+\, m_\varphi\bigg[\Big(-m^2\transv{p}{}^2 \transv{p'}{}{}^2
+2 \transv{p}{}^2 \transv{p'}{}{}^2 (\transv{p}{}\cdot\transv{p'}{})
+3m^2(\transv{p}{}\cdot\transv{p'}{})^2 \Big)V^{(\rm cf)}
\nonumber\\ &&\quad
+\, 2\Big(-m^2\transv{p}{}^2 \transv{p'}{}{}^2
-4\transv{p}{}^2\transv{p'}{}{}^2(\transv{p}{}\cdot\transv{p'}{})
+3m^2(\transv{p}{}\cdot\transv{p'}{})^2
 \Big)V^{(\rm 1g)}\bigg]\widetilde g_3 \Bigg\} \,,
\label{axial-bsw-component-3}
\ea
%and   %%%%%%%%%%%%%%%%%%%%%%%%%%%%%%%%%%%%%%%%%%%%%%%%%%%%%%%%%%%%%%%
\ba
\widetilde g_5 &=&
{1\over 2\omega m_\varphi\transv{p}{}{}^2(m_\varphi^2-4\omega^2)}
\int {d^3\transv{p'}{}\over (2\pi)^3}\Bigg\{
2m(-\transv{p}{}{}^2\transv{p'}{}{}^2
+(\transv{p}{}\cdot\transv{p'}{})^2)V^{(\rm cf)} \widetilde g_4
\nonumber\\[5pt] &&\quad
+\, 2m_\varphi^3\transv{p}{}^2 \,(V^{(\rm cf)}+2V^{(\rm 1g)})\widetilde g_7
%\nonumber\\[5pt] &&\quad
- 4m_\varphi \transv{p}{}^2 \Big[(m^2+\transv{p}{}\cdot\transv{p'}{}) V^{(\rm cf)}
-2(\transv{p}{}\cdot\transv{p'}{}) V^{(\rm 1g)}\Big]\widetilde g_5
\nonumber\\[5pt] &&\quad
+\, m_\varphi\Big[\transv{p}{}^2 \transv{p'}{}{}^2
- (\transv{p}{}\cdot\transv{p'}{})^2
\Big](V^{(\rm cf)}+2V^{(\rm 1g)}) \widetilde g_3 \Bigg\} \,,
\label{axial-bsw-component-4}
\ea
where $\omega^2=m^2-\transv{p}{}{}^2\equiv m^2+|\transv{{\bf p}}{}|^2$\,.
After carrying out the azimuthal integration of the three-momentum
$\transv{{\bf p}'}{}$, we have
\ba
\widetilde g_a^{} =
{-1\over 2\omega|\transv{{\bf p}}{}|^4(m_\varphi^2-4\omega^2)}
\sum_b\int d |\transv{{\bf p}'}{}| \Big[
B_{a b}(|\transv{{\bf p}}{}|, |\transv{{\bf p}'}{}|)
\,\widetilde g_b^{} (|\transv{{\bf p}'}{}|)
- C_{a b}(|\transv{{\bf p}}{}|, |\transv{{\bf p}'}{}|)
\,\widetilde g_b^{} (|\transv{{\bf p}}{}|)\Big]\,,
\ea
where $B_{a b}(|\transv{{\bf p}}{}|, |\transv{{\bf p}'}{}|)$
and $C_{a b}(|\transv{{\bf p}}{}|, |\transv{{\bf p}'}{}|)$
($a, b=3,4,5,7$) are defined in Appendix \ref{sect-B}\,. Using the Gaussian
quadrature method to discretize the integral equations, we obtain the
following linear equations:
\ba
\widetilde g_7^{}&=&U_3\, \widetilde g_3^{}+U_4\,\widetilde g_4^{}
+U_5\,\widetilde g_5^{}+U_7\, \widetilde g_7^{}\,,
\nonumber\\
0&=&R_3\, \widetilde g_3^{}+R_4\,\widetilde g_4^{}
+R_5\,\widetilde g_5^{}+R_7\, \widetilde g_7^{}\,,
\nonumber\\
0&=&S_3\, \widetilde g_3^{}+S_4\,\widetilde g_4^{}
+S_5\,\widetilde g_5^{}+S_7 \,\widetilde g_7^{}\,,
\nonumber\\
0&=&T_3 \,\widetilde g_3^{}+T_4\,\widetilde g_4^{}
+T_5\,\widetilde g_5^{}+T_7\, \widetilde g_7^{}\,,
\ea
where $U_i$, $R_i$, $S_i$, $T_i$ ($i=1,\dots,4$) can be read off from Eqs.
(\ref{axial-bsw-component-1}), (\ref{axial-bsw-component-2}),
(\ref{axial-bsw-component-3}), and (\ref{axial-bsw-component-4}), respectively.
After some algebras we have the following eigenvalue equation for
$\widetilde g_7^{}$:
\ba
\widetilde g_7^{}&=& U_7\,\widetilde g_7^{}
-U_3\,\big[T_{45}^{-1}T_{43}-S_{45}^{-1}S_{43}\big]^{-1}
\big[T_{45}^{-1}T_{47}-S_{45}^{-1}S_{47}\big]\,\widetilde g_7^{}
\nonumber\\&&\qquad
-\,U_4\,\big[T_{53}^{-1}T_{54}-S_{53}^{-1}S_{54}\big]^{-1}
\big[T_{53}^{-1}T_{57}-S_{53}^{-1}S_{57}\big]\,\widetilde g_7^{}
\nonumber\\&&\qquad
-\,U_5\,\big[T_{34}^{-1}T_{35}-S_{34}^{-1}S_{35}\big]^{-1}
\big[T_{34}^{-1}T_{37}-S_{34}^{-1}S_{37}\big]\,\widetilde g_7^{}\,,
\label{axial-eigen-eq}
\ea
where, for convenience, we have defined
\ba
T_{i j}=T_i^{-1}T_j - R_i^{-1}R_j\,,\qquad S_{i j}=S_i^{-1}S_j - R_i^{-1}R_j\,.
\ea
After $\tilde g_7$ is solved out from Eq.(\ref{axial-eigen-eq}), we can 
obtain $\tilde g_{3,4,5}$ by the following equations
\ba
\widetilde g_3^{}&=&-\,\big[T_{45}^{-1}T_{43}-S_{45}^{-1}S_{43}\big]^{-1}
\big[T_{45}^{-1}T_{47}-S_{45}^{-1}S_{47}\big]\,\widetilde g_7^{}\,,
\\
\widetilde g_4^{}&=&-\,\big[T_{53}^{-1}T_{54}-S_{53}^{-1}S_{54}\big]^{-1}
\big[T_{53}^{-1}T_{57}-S_{53}^{-1}S_{57}\big]\,\widetilde g_7^{}\,,
\\
\widetilde g_5^{}&=&-\,\big[T_{34}^{-1}T_{35}-S_{34}^{-1}S_{35}\big]^{-1}
\big[T_{34}^{-1}T_{37}-S_{34}^{-1}S_{37}\big]\,\widetilde g_7^{}\,.
\ea

\subsection{Normalization condition for the BS wave function of the axial-vector
diquark}

The inversion of the `free' four-point function
(\ref{forward-4-point-axial-vector-diquark}) can be defined to be
\ba
I_{P_\varphi}(p,p')^{i_1i_2,j_2j_1}_{\alpha_1\alpha_2,\beta_2\beta_1}
= {1\over 2}\delta^{i_1j_1}\delta^{i_2j_2}(2\pi)^4\delta^4(p-p')
[S(p_1)\gamma^0]^{-1}_{\alpha_1\beta_1}
[S(p_2)\gamma^0]^{-1}_{\alpha_2\beta_2}\,,
\ea
%
%
\iffalse
which satisfies
\ba
&&\int {d^4 k\over (2\pi)^4}
I_{P_\varphi}(p,k)^{i_1i_2i_2'i_1'}_{\alpha_1\alpha_2\gamma_2\gamma_1}
\widetilde S_{P_\varphi}^{(0)}(k,p')^{i_1'i_2'j_2j_1}_{\gamma_1\gamma_2\beta_2\beta_1}
\nonumber\\&&\qquad
= {1\over 2}(2\pi)^4
\Big[\delta^4(p-p')\delta_{\alpha_1\beta_1}\delta_{\alpha_2\beta_2}
\delta^{i_1j_1}\delta^{i_2j_2}
- \delta^4(p+p')\delta_{\alpha_1\beta_2}\delta_{\alpha_2\beta_1}
\delta^{i_1j_2}\delta^{i_2j_1}
\Big]\,,
\nonumber
\ea
and
\ba
&&\int {d^4 k\over (2\pi)^4}
\widetilde S_{P_\varphi}^{(0)}(p,k)^{i_1i_2i_2'i_1'}_{\alpha_1\alpha_2\gamma_2\gamma_1}
I_{P_\varphi}(k,p')^{i_1'i_2'j_2j_1}_{\gamma_1\gamma_2\beta_2\beta_1}
\nonumber\\&&\qquad
= {1\over 2}(2\pi)^4
\Big[\delta^4(p-p')\delta_{\alpha_1\beta_1}\delta_{\alpha_2\beta_2}
\delta^{i_1j_1}\delta^{i_2j_2}
- \delta^4(p+p')\delta_{\alpha_1\beta_2}\delta_{\alpha_2\beta_1}
\delta^{i_1j_2}\delta^{i_2j_1}
\Big]\,.
\nonumber
\ea
\fi
%
%
From this equation and using the fact
$\chiM{P_\varphi}{(r)}(p)_{\alpha\beta}=\chiM{P_\varphi}{(r)}(-p)_{\beta\alpha}$
we arrive at the normalization condition for the BS wave
function (see Eq. (\ref{diquark-normalization-condition})),
\ba
&&{1\over 24}\int {d^4p\over(2\pi)^4}\,{\rm Tr}\,
\bigg\{{\cal C}\gamma^0 \,\ovlchiM{P_\varphi}{(r)}(p)^{\rm T}\,\gamma^0
{\not\!\xi}\, \tildechiM{P_\varphi}{(r)}(p)^{\rm T} S(-p_1)^{-1}
\nonumber\\&&\qquad\qquad\qquad
-\, {\cal C}\gamma^0\, \ovlchiM{P_\varphi}{(r)}(p)^{\rm T}\, \gamma^0
S(p_2)^{-1}\tildechiM{P_\varphi}{(r)}(p)^{\rm T} {\not\!\xi}\, \bigg\}
= 1\,.
\ea
Similar to the case for the scalar diquark, we define
\ba
\label{axial-diquark-factorized}
\tildechiM{P_\varphi}{(r)}(p)^{\rm T}
&\equiv&i S(p_2)\tildetildechiM{P_\varphi}{(r)}(\transv{p}{}) S(-p_1)\,,
\\
{\cal C}\gamma^0 \,\ovlchiM{P_\varphi}{(r)}(p)^{\rm T}\,\gamma^0
&\equiv& i S(-p_1)\tildetildechiM{P_\varphi}{(r)}(\transv{p}{})^{(\rm c)} S(p_2)
\,,
\label{axial-diquark-factorized-cc}
\ea
where
\ba
\tildetildechiM{P_\varphi}{(r)}(\transv{p}{})&=&
\varepsilon_\rho^{(r)}(P_\varphi)
\bigg\{\transv{p}{}^\rho \, {\widetilde g_1^{}{}'\over m_\varphi}
+ \transv{p}{\mu} P_{\varphi,\nu}\gamma^{\rho\mu\nu}
{\widetilde g_5^{}{}'\over m_\varphi^2} + \gamma^\rho \widetilde g_7^{}{}'
+ \transv{p}{}^\rho \transv{{\not\!p}}{} {\widetilde g_3^{}{}'\over m_\varphi^2}
\nonumber\\[5pt]&&\qquad
-\,i g^{\rho\mu}P_\varphi^\nu {\widetilde g_6^{}{}'\over m_\varphi} \sigma_{\mu\nu}
- i\transv{p}{}^\rho \transv{p}{}^\mu P_\varphi^\nu
{\widetilde g_4^{}{}'\over m_\varphi^3}
\sigma_{\mu\nu} \bigg\}\,,
\\
\tildetildechiM{P_\varphi}{(r)}(\transv{p}{})^{(\rm c)}&=&
\varepsilon_\rho^{(r)}(P_\varphi)
\bigg\{-\transv{p}{}^\rho \, {\widetilde g_1^{}{}'\over m_\varphi}
+ \transv{p}{\mu} P_{\varphi,\nu}\gamma^{\rho\mu\nu}
{\widetilde g_5^{}{}'\over m_\varphi^2} - \gamma^\rho \widetilde g_7^{}{}'
- \transv{p}{}^\rho \transv{{\not\!p}}{} {\widetilde g_3^{}{}'\over m_\varphi^2}
\nonumber\\[5pt]&&\qquad
-\,i g^{\rho\mu}P_\varphi^\nu {\widetilde g_6^{}{}'\over m_\varphi} \sigma_{\mu\nu}
- i\transv{p}{}^\rho \transv{p}{}^\mu P_\varphi^\nu
{\widetilde g_4^{}{}'\over m_\varphi^3}
\sigma_{\mu\nu} \bigg\}\,,
\ea
and $\widetilde g_1^{}{}',\dots,\widetilde g_7^{}{}'$ are defined by the
following integrations:
\ba
\widetilde g_1^{}{}'(\transv{p}{})&=&\int {d^3\transv{p'}{}\over(2\pi)^3}
(-4V^{(\rm 1g)}+V^{(\rm cf)}){\transv{p}{}\cdot\transv{p'}{}\over\transv{p}{}^2}
\,\widetilde g_1^{}(\transv{p'}{})\,,
\\
\widetilde g_5^{}{}'(\transv{p}{})&=&\int {d^3\transv{p'}{}\over(2\pi)^3}
(-2V^{(\rm 1g)}+V^{(\rm cf)}){\transv{p}{}\cdot\transv{p'}{}\over\transv{p}{}^2}
\,\widetilde g_5^{}(\transv{p'}{})\,,
\\
\widetilde g_7^{}{}'(\transv{p}{})&=&\int {d^3\transv{p'}{}\over(2\pi)^3}
(2V^{(\rm 1g)}+V^{(\rm cf)})\bigg[\widetilde g_7^{}(\transv{p'}{})
+ {\,\transv{p'}{}{}^2\transv{p}{}^2 - (\transv{p}{}\cdot\transv{p'}{})^2
\over 2m_\varphi^2\,\transv{p}{}^2} \,\widetilde g_3^{}(\transv{p'}{})\bigg]\,,
\\
\widetilde g_6^{}{}'(\transv{p}{})&=&\int {d^3\transv{p'}{}\over(2\pi)^3}
V^{(\rm cf)}\bigg[\widetilde g_6^{}(\transv{p'}{})
+ {\transv{p'}{}{}^2\transv{p}{}^2 - (\transv{p}{}\cdot\transv{p'}{})^2
\over 2m_\varphi^2\,\transv{p}{}^2} \,\widetilde g_4^{}(\transv{p'}{})\bigg]\,,
\\
\widetilde g_3^{}{}'(\transv{p}{})
&=&{1\over 2}\int {d^3\transv{p'}{}\over(2\pi)^3}\,
{3(\transv{p}{}\cdot\transv{p'}{})^2 - \transv{p}{}^2\transv{p'}{}{}^2
\over\transv{p}{}^4}\,
(2V^{(\rm 1g)}+V^{(\rm cf)})\,\widetilde g_3^{}(\transv{p'}{})\,,
\\
\widetilde g_4^{}{}'(\transv{p}{})
&=&{1\over 2}\int {d^3\transv{p'}{}\over(2\pi)^3}\,
{3(\transv{p}{}\cdot\transv{p'}{})^2 - \transv{p}{}^2\transv{p'}{}{}^2
\over\transv{p}{}^4}\,
V^{(\rm cf)}\,\widetilde g_4^{}(\transv{p'}{})\,.
\ea
Then the normalization condition can be written as
\ba
&&-\,{1\over 24}\int {d^4p\over(2\pi)^4}\,{\rm Tr}\,
\bigg\{S(-p_1)\,\tildetildechiM{P_\phi}{(r)}(\transv{p}{})^{(\rm c)}\,
S(p_2){\not\!\xi}\,S(p_2)\tildetildechiM{P_\phi}{(r)}(\transv{p}{})
\nonumber\\&&\qquad\qquad\qquad
-\, S(-p_1)\,\tildetildechiM{P_\phi}{(r)}(\transv{p}{})^{(\rm c)}\,
S(p_2)\,\tildetildechiM{P_\phi}{(r)}(\transv{p}{})\,S(-p_1){\not\!\xi} \bigg\}
= 1\,,
\ea
which is similar to the case for the scalar diquark. After integrating out
the longitudinal momentum $\longit{p}$ and carrying out the trace calculation,
we have the following one-dimensional integral equation:
\ba
&&{2E_\varphi\over 3 m_\varphi^5}\int{|\transv{{\bf p}}{}|^2d |\transv{{\bf p}}{}|
\over 2\pi^2\omega(m_\varphi^2-4\omega^2)^2}
\bigg\{ 2m_\varphi^3 |\transv{{\bf p}}{}|^4 \widetilde g_1^{}{}'{}^2
+ 4m_\varphi^3 |\transv{{\bf p}}{}|^4 \widetilde g_5^{}{}'{}^2
+ 2m_\varphi\omega^2 |\transv{{\bf p}}{}|^4 \widetilde g_4^{}{}'{}^2
\nonumber\\&&\qquad
+ 2m^2 m_\varphi |\transv{{\bf p}}{}|^4 \widetilde g_3^{}{}'{}^2
+ 2m_\varphi^5(2m^2+\omega^2)\widetilde g_6^{}{}'{}^2
+ 2m_\varphi^5(m^2+2\omega^2)\widetilde g_7^{}{}'{}^2
\nonumber\\&&\qquad
+ m_\varphi|\transv{{\bf p}}{}|^4(m_\varphi^2+4\omega^2)
\widetilde g_1^{}{}' \widetilde g_4^{}{}'
- m |\transv{{\bf p}}{}|^4(m_\varphi^2+4\omega^2)
\widetilde g_3^{}{}' \widetilde g_4^{}{}'
- 4m m_\varphi^2 |\transv{{\bf p}}{}|^4 \widetilde g_1^{}{}' \widetilde g_3^{}{}'
\nonumber\\&&\qquad
- 4m^2m_\varphi^3 |\transv{{\bf p}}{}|^2 \widetilde g_7^{}{}' \widetilde g_3^{}{}'
+ m m_\varphi^2 |\transv{{\bf p}}{}|^2(m_\varphi^2+4\omega^2)
\widetilde g_7^{}{}' \widetilde g_4^{}{}'
+ 2m_\varphi^3 |\transv{{\bf p}}{}|^2(m_\varphi^2+4\omega^2)
\widetilde g_5^{}{}' \widetilde g_7^{}{}'
\nonumber\\&&\qquad
+ 4m m_\varphi^4 |\transv{{\bf p}}{}|^2 \widetilde g_1^{}{}' \widetilde g_7^{}{}'
- 3m m_\varphi^4(m_\varphi^2+4\omega^2)\widetilde g_6^{}{}' \widetilde g_7^{}{}'
- m_\varphi^3 |\transv{{\bf p}}{}|^2(m_\varphi^2+4\omega^2)
\widetilde g_1^{}{}' \widetilde g_6^{}{}'
\nonumber\\&&\qquad
+ m m_\varphi^2 |\transv{{\bf p}}{}|^2(m_\varphi^2+4\omega^2)
\widetilde g_6^{}{}' \widetilde g_3^{}{}'
- 8m m_\varphi^4 |\transv{{\bf p}}{}|^2 \widetilde g_5^{}{}' \widetilde g_6^{}{}'
- 4 m_\varphi^3\omega^2 |\transv{{\bf p}}{}|^2
\widetilde g_6^{}{}' \widetilde g_4^{}{}'
\bigg\} = 1\,,
\label{axial-diquark-normalization-eq}
\ea
where $E_\varphi=P_\varphi\cdot\xi=P_\varphi^0$ is the energy of the axial-vector
diquark $\varphi$.

\section{The effective interaction of diquarks and the pion}
\label{sect: interaction-of-qq-and-pion}

As pointed out in the introduction, to study the decays of baryons in the
diquark picture, we should first calculate the matrix element
$\langle \pi(q) |{\rm T}\phi^i(x)\varphi^i(y)|0\rangle$\,. This can be
determined by the effective low-energy interaction of diquarks and the pion.
We will calculate the effective coupling constant in this section by
presenting the decay amplitude of a process, where the axial-vector diquark
decays into the scalar diquark and a very soft pion, in terms of BS wave
functions of diquarks $\phi$ and $\varphi$\,.

Let us first introduce the effective interaction vertex of diquarks and the
pseudo-Goldstone-boson pion:
\ba
{\cal L}_{\pi\phi\varphi}=G_{\pi\phi\varphi}^{}\sum_{b,i}\phi^i\partial^\mu\pi^b
\,\ovl\varphi_\mu^{\,b,i}+\hbox{h.c.}\,,
\label{pi-phi-varphi-effective-lagrangian}
\ea
where $i$ is the index of the (anti-)fundamental representation of the color
group $SU(3)_c$, $\pi^b(x)$ is $\pi$ meson field with
$b$ $(=\pm,0)$ being the isospin index. Then the decay amplitude can be
calculated out to the lowest order,
\ba
\langle\phi^i(P_\phi) \pi^a(q)| \varphi^{a,i}(P_\varphi),r\rangle
=(2\pi)^4\delta^4(q+P_\phi-P_\varphi)
{-3G_{\pi\phi\varphi}^{}\over\sqrt{2E_\pi}\sqrt{2E_\phi}\sqrt{2E_\varphi}}\,
q^\mu \varepsilon_\mu^{(r)}(P_\varphi)\,,
\label{diquark-decay-lowest-order}
\ea
where the summation over the repeated color index $i$ is assumed and
no summation is assumed for the repeated isospin index $a$.
On the other hand, we can present this decay amplitude in terms of the
following transition amplitude with the aid of the partial conservation of
the axial currents (PCAC)\cite{PCAC},
$\partial^\mu A_\mu^a(x)\approx\sqrt{2} m_\pi^2 f_\pi \pi^a(x)$:
\ba
\langle\phi^i(P_\phi) \pi^a(q)| \varphi^{a,i}(P_\varphi),r\rangle
&\makebox[0pt]{=}&\int d^4x\, e^{iqx}\, i(\square_x+m_\pi^2)
\langle\phi^i(P_\phi) | \pi^a(x)^\dag | \varphi^{a,i}(P_\varphi),r\rangle
\nonumber\\&\makebox[0pt]{=}&
-\,{q^2-m_\pi^2\over m_\pi^2 \sqrt{2E_\pi}}\,{q^\mu\over \sqrt{2} f_\pi}
\langle\phi^i(P_\phi) | A_\mu^a(q)^\dag | \varphi^{a,i}(P_\varphi),r\rangle \,,
\label{diquark-decay-PCAC}
\ea
where $f_\pi\approx 93.3$\,MeV is the pion decay constant, and the
axial-vector current is $A_\mu^-(z)=\ovl \psi_d\gamma_5\gamma_\mu \psi_u(z)$\,.
Furthermore, momentum conservation and Lorentz invariance lead to
\ba
q^\mu \langle\phi^i(P_\phi) | A_\mu^-(q) | \varphi^{+,i}(P_\varphi),r\rangle
\sqrt{2E_\phi 2E_\varphi}
= P_\phi^{\rm t}\cdot\varepsilon^{(r)}\,
(2\pi)^4\delta^4(q+P_\phi-P_\varphi)\,{\cal A}(q^2)\,,
\label{diquark-transition-amplitude-lorentz-structure}
\ea
where $q=P_\varphi-P_\phi$ and $P_\varphi\cdot\varepsilon^{(r)}=0$ have been used.
The factors $\sqrt{2E_\phi}$ and $\sqrt{2E_\varphi}$ are introduced in Eq.
(\ref{diquark-transition-amplitude-lorentz-structure}) for latter convenience
      \footnote{
        In fact, by including these energy factors in the definition of the
        transition amplitude we make ${\cal A}$ be a Lorentz invariant quantity
        since we have used the normalization convention for the single
        particle state,
        $\langle{\bf p}|{\bf p'}\rangle=(2\pi)^2\delta^3({\bf p}-{\bf p'})$\,.
      }.
This can been seen from the normalization equations
(\ref{scalar-diquark-normalization-eq}) and
(\ref{axial-diquark-normalization-eq}): with these energy factors absorbed into
the BS wave functions of the diquarks, we shall only normalize
$\sqrt{2E_\phi}\,\chiM{P_\phi}{}$ and $\sqrt{2E_\varphi}\,\chiM{P_\varphi}{(r)}$,
which are Lorentz covariant quantities, instead of $\chiM{P_\phi}{}$ and
$\chiM{P_\varphi}{(r)}$\,, respectively.

Combining Eqs. (\ref{diquark-decay-lowest-order}), (\ref{diquark-decay-PCAC}),
and (\ref{diquark-transition-amplitude-lorentz-structure}) gives
\ba
{m_\pi^2\over q^2-m_\pi^2}\,G_{\pi\phi\varphi}^{}
={-1\over 3\sqrt{2} f_\pi}\,{\cal A}(q^2)\,.
\ea
This equation shows that the amplitude ${\cal A}$ must develop a pole at
$q^2=m_\pi^2$\,. As usual, we can decompose this amplitude into two terms:
one has the desired pole, the other is regular:
\ba
{\cal A}(q^2)={\cal B}_1(q^2)+{q^2\over q^2-m_\pi^2}\,{\cal B}_2(q^2)\,,
\ea
where ${\cal B}_1$ and ${\cal B}_2$ are smooth functions without extra poles at
$q^2=m_\pi^2$\,, and their dependence on $q^2$ are expected to be very weak
and hence are nearly constants when $q^2$ is small enough, e.g.
$q^2\in (0,m_\pi^2)$\,. With these expectations in mind, we have the following
Goldberger-Treiman-like relations:
\ba
{\cal B}_1(q^2)\approx 3\sqrt{2}f_\pi G_{\pi\phi\varphi}\,,\qquad
{\cal B}_2(q^2)\approx -3\sqrt{2}f_\pi G_{\pi\phi\varphi}\,.
\ea
Then, the effective coupling constant can be calculated approximately by
\ba
G_{\pi\phi\varphi}^{} \approx {1\over 3\sqrt{2} f_\pi}\,{\cal A}(0)\,.
\label{effective-coupling-constant}
\ea
The transition amplitude ${\cal A}(q^2)$ at $q^2=0$ is regular and it is this
value that will be calculated in the following.

To calculate ${\cal A}$\,, one can choose a specific polarization vector.
Choosing $\varepsilon^{(r)}=P_\phi^{\rm t}$ and using
Eq. (\ref{diquark-transition-amplitude-lorentz-structure}), we have
\ba
(2\pi)^4\delta^4(q+P_\phi-P_\varphi)\,{\cal A}(q^2)
= {q^\mu \langle\phi^i(P_\phi) |A_\mu^-(q)| \varphi^{+,i}(P_\varphi),r\rangle
\big|_{\varepsilon^{(r)}=P_\phi^{\rm t}} \over (P_\phi^{\rm t})^2 }
\sqrt{2E_\phi 2E_\varphi}\,.
\label{diquark-transition-amplitude-special-polarization}
\ea
Let us now turn to the calculation of the matrix element
$\langle\phi^i(P_\phi) |A_\mu^-(q)| \varphi^{+,i}(P_\varphi),r\rangle$\,, which
can be represented in terms of the BS wave functions of the diquarks $\phi$
and $\varphi$ (using Eqs. (\ref{scalar-diquark-factorized-cc}) and
(\ref{axial-diquark-factorized})),
\ba
&&\langle\phi^i(P_\phi) | A_\mu^-(q) | \varphi^{+,i}(P_\varphi),r\rangle
= {1\over 12}(2\pi)^4\delta^4(q+P_\phi-P_\varphi)
\int {d^4p d^4p'\over (2\pi)^8} (2\pi)^4\delta^4(p_2-p_2')
\nonumber\\&&\qquad\qquad\times\,
{\rm Tr}\,\bigg\{ S(-p_1')\tildetildechiM{P_\phi}{}(\transv{p'}{})^{(c)}
S(p_2)\tildetildechiM{P_\varphi}{(r)}(\transv{p}{})S(-p_1)\gamma_\mu\gamma_5
\bigg\}
\nonumber\\[5pt]
&&= {1\over 12}(2\pi)^4\delta^4(q+P_\phi-P_\varphi)
\int {d^4p\over (2\pi)^4}
%\nonumber\\&&\quad\times\,
{\rm Tr}\,\bigg\{ S(-p_1')\tildetildechiM{P_\phi}{}(\transv{p'}{})^{(c)}
S(p_2)\tildetildechiM{P_\varphi}{(r)}(\transv{p}{})S(-p_1)\gamma_\mu\gamma_5
\bigg\}\,,
\nonumber\\
\label{diquark-transition-eq1}
\ea
where $p_1=P_\varphi/2+p$\,, $p_2=P_\varphi/2-p$\,,
$p_1'=P_\phi/2+p'$\,, $p_2'=P_\phi/2-p'$\,, and the delta function
$\delta^4(p_2-p_2')$ imposes the following constraints upon the variables
in the integral:
\ba
\longit{p'}&=&
{-|\transv{{\bf p}}{}||P_\phi^{\rm t}|\cos\theta +m_\phi^2 - P_\phi\cdot P_\varphi
+\longit{p} P_\phi^{\ell}\over m_\phi}\,,
\\
\transv{p'}{}{}^2 &=& -|\transv{{\bf p}}{}|^2
- |\transv{{\bf p}}{}||P_\phi^{\rm t}|\cos\theta
- |P_\phi^{\rm t}|^2/4+(\longit{p}+P_\phi^\ell/2-m_\varphi/2)^2-\longit{p'}{}^2\,,
\\
\transv{p'}{}\cdot P_\phi^{\rm t} &=&
- |\transv{{\bf p}}{}||P_\phi^{\rm t}|\cos\theta
- (1/2-\longit{p'}/m_\phi)|P_\phi^{\rm t}|^2\,,
\\
\transv{p'}{}\cdot\transv{p}{} &=& - |\transv{{\bf p}}{}|^2
- (1/2-\longit{p'}/m_\phi)|\transv{{\bf p}}{}||P_\phi^{\rm t}|\cos\theta\,,
\\
\transv{p'}{}\cdot P_\varphi&=& \longit{p} m_\varphi - m_\varphi^2/2
+ (1/2-\longit{p'}/m_\phi)P_\phi\cdot P_\varphi\,,
\ea
where $\longit{p}=p\cdot P_\varphi/m_\varphi$\,,
$\longit{p'}=p'\cdot P_\phi/m_\phi$,
$\transv{p}{}=p-{\longit{p}\over m_\varphi}P_\varphi$\,,
$\transv{p'}{}=p'-{\longit{p'}\over m_\phi}P_\phi$, and $\cos\theta$ is the
azimuthal angle between $\transv{p}{}$ and $P_\phi^{\rm t}$\,.
By momentum conservation, $ |P_\phi^{\rm t}|$ and $ P_\phi\cdot P_\varphi $ can be 
written as:
\ba
P_\phi\cdot P_\varphi = {m_\phi^2+m_\varphi^2-q^2\over 2}
\equiv P_\phi^\ell m_\varphi\,,\qquad
|P_\phi^{\rm t}|^2 = - m_\phi^2 + (P_\phi^\ell)^2\,,
\ea
where we have defined
$P_\phi^{\rm t}=P_\phi-{P_\phi^\ell\over m_\varphi}P_\varphi$\,.

To integrate out the longitudinal momentum $\longit{p}$, we decompose
the propagators $S(-p_1')$\,, $S(p_2)$, and $S(-p_1)$ into
\ba
S(-p_1') &\makebox[0pt]{=}& {({\not\!p}_1'-m)\over 2\omega_q}\bigg[
{-i\over \longit{p}+P_\phi^\ell-m_\varphi/2-\omega_q+i\epsilon}
- {-i\over \longit{p}+P_\phi^\ell-m_\varphi/2+\omega_q-i\epsilon} \bigg]\,,\quad
\\[5pt]
S(p_2) &\makebox[0pt]{=}& {-i\over \longit{p}-m_\phi/2+\omega_p-i\epsilon}
\widetilde \Lambda_2^+(\transv{p}{})
+ {-i\over \longit{p}-m_\varphi/2-\omega_p+i\epsilon}
\widetilde \Lambda_2^-(\transv{p}{})\,,
\\[5pt]
S(-p_1) &\makebox[0pt]{=}& {-i\over \longit{p}+m_\phi/2-\omega_p+i\epsilon}
\widetilde \Lambda_1^+(\transv{p}{})
+ {-i\over \longit{p}+m_\varphi/2+\omega_p-i\epsilon}
\widetilde \Lambda_1^-(\transv{p}{})\,,
\ea
where $p_1'=\transv{p}{}+{\longit{p}\over m_\varphi}P_\varphi+P_\phi^{\rm t}
+({P_\phi^\ell\over m_\varphi}-{1\over 2})P_\varphi$\,,
$\omega_q=\sqrt{m^2+|\transv{{\bf p}}{}|^2+|P_\phi^{\rm t}|^2
+ 2|\transv{{\bf p}}{}||P_\phi^{\rm t}|\cos\theta}$\,,
$\omega_p=\sqrt{m^2+|\transv{{\bf p}}{}|^2}$\,, $m$ is the mass of the
constituent quark within the diquarks $\varphi$ and $\phi$\,. The modified
projection operators are defined by
$\widetilde \Lambda_2^\pm={\myslash{P_\varphi}\over m_\varphi}\Lambda_2^\pm$
and
$\widetilde \Lambda_1^\pm={\cal C}^{-1}(\Lambda_1^\pm)^{\rm T}{\cal C}
{\myslash{P_\varphi}\over m_\varphi}$ and can be written out explicitly as
\ba
\widetilde \Lambda_2^\pm(\transv{p}{}) = {\myslash{P_\varphi\over 2m_\varphi}}
\pm {-{\not\!\transv{p}{}}+m\over 2\omega_p}
\equiv \widetilde \Lambda_1^\mp(\transv{p}{})\,.
\ea
Carrying out the integration over the longitudinal momentum $\longit{p}$\,,
the transition amplitude (\ref{diquark-transition-eq1}) becomes
\ba
&&\langle\phi^i(P_\phi) |A_\mu^-(q)| \varphi^{+,i}(P_\varphi),r\rangle
= {1\over 12}(2\pi)^4\delta^4(q+P_\phi-P_\varphi)
\int {d^3\transv{p}{}\over (2\pi)^3}
\nonumber\\[5pt]&&~\times\,
\Bigg\{
{{\rm Tr}\,(++)\big|_{\longit{p}=m_\varphi/2-\omega_p}
\over (2\omega_p-m_\varphi)(P_\phi^\ell-\omega_p-\omega_q)}
+{{\rm Tr}\,(+-)\big|_{\longit{p}=m_\varphi/2+\omega_q-P_\phi^\ell}
\over (\omega_p+\omega_q-P_\phi^\ell)(m_\varphi+\omega_p+\omega_q-P_\phi^\ell)}
\nonumber\\[5pt]&&\qquad
-\,{{\rm Tr}\,(--)\big|_{\longit{p}=-m_\varphi/2-\omega_p}
\over (2\omega_p+m_\varphi)(m_\varphi+\omega_p+\omega_q-P_\phi^\ell)}
+{ {\rm Tr}\,(++)\big|_{\longit{p}=-m_\varphi/2+\omega_p}
\over (2\omega_p-m_\varphi)(m_\varphi-\omega_p-\omega_q-P_\phi^\ell)}
\nonumber\\[5pt]&&\qquad
-\,{{\rm Tr}\,(-+)\big|_{\longit{p}=m_\varphi/2-\omega_q-P_\phi^\ell}
\over (\omega_p+\omega_q+P_\phi^\ell)(m_\varphi-\omega_p-\omega_q-P_\phi^\ell)}
-{{\rm Tr}\,(--)\big|_{\longit{p}=m_\varphi/2+\omega_p}
\over (\omega_p+\omega_q+P_\phi^\ell)(m_\varphi+2\omega_p)}
\Bigg\}\,,\qquad
\label{diquark-transition-amplitude-final-expression}
\ea
where ${\rm Tr}\,(\pm\pm)$ are abbreviations for the following expressions:
\ba
{\rm Tr}\,(\pm\pm)={\rm Tr}\,\bigg\{
{({\not\!p}_1'-m) \over 2\omega_q}\,
\tildetildechiM{P_\phi}{}(\transv{p'}{})^{(c)}\,\,
\widetilde\Lambda_2^\pm\,\, \tildetildechiM{P_\varphi}{(r)}(\transv{p}{})\,\,
\widetilde\Lambda_1^\pm \,\,\gamma_\mu\gamma_5 \bigg\}\,.
\ea
The traces % (calculated by the software:
%``\,MATHEMATICA 5.2\,''+``\,FeynCalc 4\,'')
are too lengthy to be expressed here (and so are the transition amplitudes).
Therefore, we will give only the numerical results in the following.

\bigskip\bigskip\noindent
{\it Numerical results}
\bigskip

\noindent
Motivated by the studies of mesons in the potential model
\cite{egkly-prd17-3090} and in the BS formalism \cite{meson-in-BS}, we take
the parameter $\kappa=0.2$\,GeV$^2$ in the effective potential
(\ref{cornell-potential}) and vary the effective coupling constant $\alpha_s$
to find the solutions of the BS equations. For each value of the diquark
mass, $\alpha_s$ takes a certain value. On the other hand, from the analysis
of the spectrum of the heavy baryons, Ref. \cite{jaffe-diquark-report} shows
that the mass of the scalar diquark is related to that of the axial-vector
diquark by $m_\varphi-m_\phi\approx 0.210\hbox{\,GeV}$\,. We find that for the
scalar diquark, $(m_\phi(\hbox{GeV}),\alpha_s)=(0.70,0.590)$,
$(0.75,0.570)$, $(0.80,0.555)$, while for the axial-vector diquark,
$(m_\varphi(\hbox{GeV}),\alpha_s)=(0.91,0.490)$, $(0.96,0.307)$,
$(1.01,0.093)$\,. With the solutions of the BS equations as the input, the
transition amplitude in Eq.
(\ref{diquark-transition-amplitude-final-expression})
can be calculated out. Then, from Eqs. (\ref{effective-coupling-constant})
and (\ref{diquark-transition-amplitude-special-polarization})
we obtain the values of ${\cal A}(0)$ and $G_{\pi\phi\varphi}$\,. The results
are listed in Table \ref{table: diquark-decay-constant}.
\begin{table}[htb]
\caption{\label{table: diquark-decay-constant}
  The transition amplitude
  ${\cal A}(q^2)$ at $q^2=0$ and the effective coupling constant
  $G_{\pi\phi\varphi}$ corresponding to various values of the mass of the scalar
  diquark $\phi$ (the mass of the axial-vector diquark is related to that of
  the scalar diquark by $m_\varphi-m_\phi\approx 0.21\hbox{\,GeV}$).
}
\center
\begin{tabular}{lrrr}
\hline\hline
 $m_\phi$\,(GeV)& 0.70 & 0.75 & 0.80 \\
\hline
 ${\cal A}$(0)\,(GeV) & 1.32 & 1.43 & $-$1.53  \\
\hline
 $G_{\pi\phi\varphi}$ & 3.35 &  3.62 & $-$3.88 \\
\hline\hline
\end{tabular}
\end{table}
It should be noted that the sign of the coupling constant $G_{\pi\phi\varphi}$ 
is an artifact in numerical calculation (which is related to the arbitrariness 
of the sign for the BS wave functions). A physically
observable quantity, e.g. the decay width discussed in the following, depends
only on the norm of the effective coupling constant $|G_{\pi\phi\varphi}|$.

At this point, we want to point out that the numerical results show that the
``$++$'' component (the sum of the first and fourth terms in Eq.
(\ref{diquark-transition-amplitude-final-expression})) gives the most important
contribution and the ``$+-$'' and ``$-+$'' components give less important
contribution ($\leq 15\%$) to the total transition amplitude. The
``$--$'' component (the sum of the third and sixth terms in Eq.
(\ref{diquark-transition-amplitude-final-expression})) gives very small
contribution to the total diquark-transition amplitude.

\section{Strong decays of heavy baryons}
\label{sect-baryon-decay}

In this section, we will turn to the calculation of the decay widths
of the processes $\Sigma_Q^{(*)}\to\Lambda_Q +\pi$\,. As pointed out in the
introduction, the heavy baryons $\Lambda_Q$ and $\Sigma_Q^{(*)}$ are regarded
as bound states of the heavy quark $Q$ and the diquarks, which are composed of
two light quarks. In this picture, we can express the decay amplitudes
in terms of the BS wave functions of $\Lambda_Q$ and $\Sigma_Q^{(*)}$
(taking $\Sigma_Q^{++}$ as an example),
\ba
&&\left\langle\Lambda_Q^+(P_{\!\Lambda}^{}) \pi^+(q)\right|
\left.\Sigma_Q^{++}(P_\Sigma)\right\rangle
\nonumber\\ &&\hspace{1cm}
=\int d^4(x_1 x_2 y_1 y_2 u v)\,
\ovlchiM{P_{\!\Lambda}^{}}{}(x_2,x_1)S_Q(x_1-y_1)^{-1}
\chiM{P_\Sigma}{}{}_{,\lambda}(y_1,y_2)
\nonumber\\ &&\hspace{1cm}\quad\times\,
\Delta_\phi^{-1}(x_2-u)\Delta_\varphi^{-1,\nu\lambda}(v-y_2)
\sum_{i}\langle \pi^+(q)|{\rm T}\,\phi^i(u)\ovl\varphi^{\,i}_\nu(v)\rvac\,,
\label{matrix-element-bsw-rep}
\ea
where the superscript $i$ is the color index, $\phi^i$ is the field of the
scalar diquark and $\varphi^i_\mu$ is the field of the axial-vector diquark.
$\Delta_\phi^{-1}$ and $\Delta_\varphi^{-1,\nu\lambda}$ are the inverse of
the propagators of the scalar diquark and the axial-vector diquark,
respectively. They are defined by
$\Delta_\phi^{-1}(x,y)\Delta_\phi(y,z)=\delta^4(x-z)$ and
$\Delta_\varphi^{-1,\nu\lambda}(x,y)\Delta_{\varphi,\lambda\mu}(y,z)
=\delta_\mu^\nu\delta^4(x-z)$. $S_Q$ ($Q=c,b$) is the quark propagator. The
BS wave function of the heavy baryons are defined by
\ba
\chiM{P_{\!\Lambda}^{}}{}(x_1,x_2) &=&
\lvac {\rm T}\, \psi_Q^i(x_1) \phi^i(x_2)|P_{\!\Lambda}^{}\rangle\,,
\\
\chiM{P_\Sigma}{}{}_{,\mu}(y_1,y_2) &=&
\lvac {\rm T}\, \psi_Q^i(y_1) \varphi_\mu^i(y_2)|P_\Sigma\rangle\,.
\ea
If not explicitly pointed out, the summation is understood for repeated
color index $i$\,. The complex conjugate of the BS wave function is defined by
$\ovlchiM{}{}(x_2,x_1)=\chiM{}{}(x_1,x_2)^*\gamma^0$\,.

To simplify the analysis, we take, as usual, the propagators of the
diquarks and the heavy quark to have the forms of the free ones. Then we have
\ba
\Delta_\phi^{-1}(x,y) &=& i (\square_x +m_\phi^2)\delta^4(x-y)\,,  \\
\Delta_\varphi^{-1,\nu\lambda}(x,y) &=& i \Big[(\square_x + m_\varphi^2)g^{\nu\lambda}
+ \partial_x^\nu \partial_x^\lambda \Big]\delta^4(x-y)\,,
\\
S_Q(x,y)&=&\int {d^4p\over(2\pi)^4}\,{i\over \not{\! p}-m_Q+i\varepsilon}
e^{-ipx}\,,
\ea
where $m_Q$ is the constituent mass of the heavy quark which can be determined
by, e.g. fitting experimental data to the results of the potential model for
mesons, $m_\phi$ and $m_\varphi$ are the masses of diquarks $\phi$ and
$\varphi$\,, respectively.

The matrix element
$\langle \pi^+(q)|{\rm T}\,\phi^i(u)\ovl\varphi^{\,i}_\nu(v)\rvac$ can be
calculated out to the lowest order with the effective interaction vertex
defined in Eq. (\ref{pi-phi-varphi-effective-lagrangian}),
\ba
\langle \pi^+(q)|{\rm T}\,\phi^i(u)\ovl\varphi^{\,i}_\nu(v)\rvac
=-3G_{\pi\phi\varphi}^{}\,{q^\mu\over\sqrt{2E_\pi}}
\int d^4 z\,e^{i q z}\, \Delta_\phi(u-z)\Delta_\varphi(z-v)_{\mu\nu}\,,
\ea
where the factor 3 comes from the summation over the color index. Then
Eq. (\ref{matrix-element-bsw-rep}) becomes
\ba
&&\langle\Lambda_Q^+(P_{\!\Lambda}^{})\pi^+(q)|\Sigma_Q^{++}(P_\Sigma)\rangle
={-3G_{\pi\phi\varphi}^{}\over\sqrt{2E_\pi}}
(2\pi)^4\delta^4(q+P_\Lambda-P_\Sigma)\, q^\lambda
\nonumber\\&& \quad\times
\int {d^4p' d^4p\over (2\pi)^8} \, (2\pi)^4\delta^4(p_1'-p_1)\,
\ovlchiM{P_{\!\Lambda}^{}}{}(p')\,S_Q(p_1)^{-1}
\chiM{P_\Sigma}{}{}_{,\lambda}(p)\,,
\label{matrix-element-bsw-p-rep-part}
\ea
where $p_1$ is the momentum of the heavy quark $Q$ within the baryon
$\Sigma_Q^{(*)}$ and $p_1'$ is the momentum of the heavy quark $Q$ within the
baryon $\Lambda_Q$, $p_2$ and $p_2'$ are the momenta of the diquarks $\varphi$
and $\phi$\,, respectively. These momenta are related to the total momenta of
the bound states by the following equations:
\ba
p_1'=\lambda_1' P_{\!\Lambda}^{} + p'\,,\quad p_2'=\lambda_2' P_{\!\Lambda}^{} - p'\,,\quad
p_1 =\lambda_1  P_\Sigma  + p \,,\quad p_2 =\lambda_2  P_\Sigma  - p \,,
\ea
where the parameters $\lambda_{1,2}$ and $\lambda_{1,2}'$ are defined by
\ba
\lambda_1' = {m_Q\over m_Q+m_\phi}\,,\quad \lambda_2'={m_\phi\over m_Q+m_\phi}\,,\quad
\lambda_1  = {m_Q\over m_Q+m_\varphi}\,,\quad \lambda_2={m_\varphi\over m_Q+m_\varphi} \,.
\ea
Following Refs. \cite{guo-1} and \cite{guo-2}, we can present the BS wave
functions of the baryons as
\ba
\ovlchiM{P_{\!\Lambda}^{}}{}(p') =
i\, \ovl u_\Lambda(v')N(\transv{p'}{},\longit{p'})\Delta_\phi(-p_2') S_Q(p_1')\,,
\ea
where $u_\Lambda$ is the Dirac spinor of $\Lambda_Q$\,.
Since there is a delta-function $\delta(p_1'-p_1)$ in
Eq. (\ref{matrix-element-bsw-p-rep-part}), to the leading order in $1/m_Q$
expansion, we have
\ba
\ovlchiM{P_{\!\Lambda}^{}}{}(p') \, S_Q(p_1)^{-1} =
{i\over (\longit{p}+m_\Sigma-P_{\!\Lambda}^\ell)^2-W_{q}^2+i\varepsilon }
\ovl u_\Lambda(v')N(\transv{p'}{},\longit{p'})\,.
\label{guo-ref-1-eq-12}
\ea
Similarly, to the leading order in $1/m_Q$ expansion, we can present the BS
wave function in the baryon $\Sigma_Q^{(*)}$ as
\ba
\chiM{P_\Sigma}{\lambda}(p) &=&
{-i\over (\longit{p}+E_0 +m_\varphi+i\varepsilon)
(\longit{p}^2-W_{p}^2+i\varepsilon)} \,
M^{\lambda\mu}(\transv{p}{},\longit{p}) B^{(m)}_{\mu}(v)\,,
\label{guo-ref-2-eq-28}
\ea
where $m=1$ is for $\Sigma_Q$\,, $m=2$ for $\Sigma_Q^*$\,, $E_0$ is the
binding energy of the baryon $\Sigma_Q^{(*)}$ which will be determined in
Sect. \ref{sect: numerical-results}. For more details on discussions about the
BS wave functions of heavy baryons, the readers are referred to the original
references \cite{guo-1,guo-2}. For convenience, we also give some results in
Appendix \ref{appendix-ref-guo}. In the above equations, we have defined
the following quantities:
$W_{q}=\sqrt{-(\transv{p}{}-P_{\!\Lambda}^{\,\rm t})^2+m_\phi^2}$\,,
$W_{p}=\sqrt{-\transv{p}{}^2+m_\varphi^2}$\,,
$\longit{p'}=p'\cdot v'-\lambda_2' m_\Lambda$\,, $\longit{p}
=p\cdot v-\lambda_2 m_\Sigma$\,.
$\transv{p'}{}=p'-(p'\cdot v')v'$ and $\transv{p}{}=p-(p\cdot v)v$ are
perpendicular to the ``velocities'' of the baryons,
$v'=P_{\!\Lambda}^{}/m_\Lambda^{}$ and $v=P_\Sigma/m_\Sigma^{}$\,, respectively.
$P_{\!\Lambda}^\ell=(m_\Lambda^2+m_\Sigma^2-q^2)/(2 m_\Sigma)=m_\Lambda^{} v\cdot v'$
is a constant,
$P_{\!\Lambda}^{\rm t}=P_{\!\Lambda}^{}-{P_{\!\Lambda}^\ell\over m_\Sigma} P_\Sigma$ is
perpendicular to $P_\Sigma$\,. The polarization vectors of the baryons
$\Sigma_Q$ and $\Sigma_Q^*$ are given by
\begin{eqnarray}
B_{\mu}^{(1)}(v) = \frac{1}{\sqrt{3}}(\gamma_\mu +v_\mu)\gamma_5 u_\Sigma^{}(v),
\quad  B_{\mu}^{(2)}(v) = u_{\mu} (v),
\label{2a}
\end{eqnarray}
where $u_\Sigma^{}(v)$ is the Dirac spinor and $u_\mu (v)$ is the
Rarita-Schwinger vector spinor. $B_{\mu}^{(m)}(v)$ satisfies the following
conditions:
\begin{equation}
\rlap / v B_{\mu}^{(m)}(v)=B_{\mu}^{(m)}(v), \quad  v^\mu B_{\mu}^{(m)}(v)=0,
\quad \gamma^\mu B_{\mu}^{(2)}(v)=0.
\label{2b}
\end{equation}
The above constraints for $m=1$ can be seen from
$\rlap / v u_\Sigma(v)=u_\Sigma(v)$ while for $m=2$, they are the properties of
the Rarita-Schwinger vector spinor with spin ${3\over 2}$. The expressions of
$N(\transv{p'}{},\longit{p'})$ and $M^{\lambda\mu}(\transv{p}{},\longit{p})$
are given in Appendix \ref{appendix-ref-guo} (these expressions are
extracted from Eq. (12) in
Ref. \cite{guo-1} and Eq. (28) in Ref. \cite{guo-2}, some notations in this
paper are different from those in the original references). On the other
hand, the delta-function $\delta(p_1-p_1')$ leads to the following relations:
\ba
\transv{p}{}\cdot P_{\!\Lambda}^{\rm t}
&=& - m_\Lambda^{} |\transv{p}{}|\sqrt{\Omega^2-1}\cos\theta\,,
\\
\longit{p'}&=&\Omega\,\longit{p}-|\transv{p}{}|\sqrt{\Omega^2-1}\cos\theta
+\Omega m_\Sigma-m_\Lambda\,,
\\
\transv{p'}{}{}^2&=& -\,|\transv{p}{}|^2\sin^2\theta
-\Big[\Omega|\transv{p}{}|\cos\theta-(m_\Sigma+\longit{p})\sqrt{\Omega^2-1}
\,\Big]^2\,,
\ea
where $\Omega=v'\cdot v$\,.

The poles of $\longit{p}$ and $\longit{p'}$ are all from the prefactors
in Eqs.\,(\ref{guo-ref-1-eq-12}) and (\ref{guo-ref-2-eq-28}) since
$N(\transv{p'}{},\longit{p'})$ depends linearly on $\longit{p'}$ and
$M^{\lambda\sigma}(\transv{p}{},\longit{p})$ contains $\longit{p}$ up to second
order \cite{guo-1,guo-2}. After carrying out the integration over $\longit{p}$
in Eq. (\ref{matrix-element-bsw-p-rep-part}), we have
\ba
\langle\Lambda_Q^+(P_{\!\Lambda}^{})\pi^+(q)|\Sigma_Q^{++}(P_\Sigma)\rangle
={3i G_{\pi\phi\varphi}^{}\,\ovl u_\Lambda(v') B^{(m)}_\rho(v)
\over\sqrt{2E_\pi}\sqrt{2E_\Lambda}\sqrt{2E_\Sigma} }
{\cal C}^\rho(q^2)\,(2\pi)^4\delta^4(q+P_\Lambda-P_\Sigma)\,,
\ea
where
\ba
{\cal C}^\rho(q^2)&\makebox[0pt]{=}&\sqrt{2E_\Lambda}\sqrt{2E_\Sigma}
\int {d^3\transv{p}{}\over (2\pi)^3} \, \Bigg\{
{ q_\lambda^{}M^{\lambda\rho}(p) N(p')|_{\longit{p}=W_p}
\over 2W_p(W_p+E_0^\Sigma +m_\varphi)
\big[(W_p+m_\Sigma-P_{\!\Lambda}^\ell)^2-W_q^2\big]}
\nonumber\\[5pt]&\makebox[0pt]{}&\quad
+\, {q_\lambda^{}M^{\lambda\rho}(p) N(p')|_{\longit{p}=-m_\Sigma+W_q+P_{\!\Lambda}^\ell}
\over 2W_q(-m_\Sigma+P_{\!\Lambda}^\ell+W_q+E_0^\Sigma +m_\varphi)
\big[(-m_\Sigma+P_{\!\Lambda}^\ell+W_q)^2-W_p^2\big]}
\nonumber\\[5pt]&\makebox[0pt]{}&\quad
+\, {q_\lambda^{}M^{\lambda\rho}(p) N(p')|_{\longit{p}=-m_\varphi-E_0^\Sigma}
\over \big[(-m_\varphi-E_0^\Sigma+m_\Sigma-P_\Lambda^\ell)^2-W_q^2\big]
\big[(m_\varphi+E_0^\Sigma)^2-W_p^2\big]}
\Bigg\}\,.
\label{matrix-element-bsw-p-rep-part-2}
\ea
The tensor function $M^{\lambda\rho}$ is defined by
\ba
M^{\lambda\rho}(p)
= g^{\lambda\rho} M_1(p)
+\,\frac{v^\lambda \transv{p}{}^\rho}{m_\varphi} M_2(p)
-\,\frac{\transv{p}{}^\lambda \transv{p}{}^\rho}{m_\varphi^{2}} M_3(p)\,,
\ea
where $g^{\mu\nu}={\rm diag}\{1,-1,-1,-1\}$ is the Lorentz metric tensor,
$M_n$ $(n=1,2,3)$ are defined in
Eqs. (\ref{sigma-tensor-function-M1})-(\ref{sigma-tensor-function-M3})
in Appendix \ref{appendix-ref-guo}. When calculating the contraction, we will
use the orthogonal condition $v^\rho B^{(m)}_\rho(v)=0$. Furthermore, the
integrations of terms containing $\transv{p}{}^\rho B^{(m)}_\rho(v)$ have the
following form:
\ba
\int d^3\transv{p}{}\, \transv{p}{}^\rho B^{(m)}_\rho(v)
h(\transv{p}{},P_\Lambda^{{\rm t}})
= P_\Lambda^{{\rm t},\rho}B^{(m)}_\rho(v) g(|P_\Lambda^{{\rm t}}|)\,,
\ea
where $h$ is arbitrary smooth function and
\ba
g(|P_\Lambda^{{\rm t}}|)=\int d^3\transv{p}{}\,
{\transv{p}{}\cdot P_\Lambda^{\rm t}\over (P_\Lambda^{\rm t})^2} \,
h(\transv{p}{},P_\Lambda^{{\rm t}})\,.
\ea
Then only terms containing the factor
$P_\Lambda^{{\rm t},\rho}B^{(m)}_\rho(v)$ can appear in the final expression.
The calculation is straightforward and we have
\ba
&\makebox[0pt]{}& {\cal C}^\rho(q^2)=P_\Lambda^{{\rm t},\rho}\,{\cal D}(q^2)=
P_\Lambda^{{\rm t},\rho}\sqrt{2E_\Lambda}\sqrt{2E_\Sigma}
\int_0^\infty {|\transv{{\bf p}}{}|^2 d |\transv{{\bf p}}{}|\over 4\pi^2}
\int_{-1}^{+1} d\cos\theta\, {1\over m_\varphi^2|P_{\!\Lambda}^{\rm t}|}
\nonumber\\[10pt]&\makebox[0pt]{}&\times\, \Bigg\{
N_a\,{(m_\Sigma-m_\Lambda\Omega)m_\varphi|\transv{p}{}|\cos\theta M_{2a}
-|P_{\!\Lambda}^{\rm t}| (m_\varphi^2 M_{1a}+|\transv{p}{}|^2\cos^2\theta M_{3a})
\over
2W_p (W_p+E_0^\Sigma +m_\varphi)\big[(W_p+m_\Sigma-P_{\!\Lambda}^\ell)^2-W_q^2\big] }
\nonumber\\[10pt]&\makebox[0pt]{}&\quad
+\,N_b\,{(m_\Sigma-m_\Lambda\Omega)m_\varphi|\transv{p}{}|\cos\theta M_{2b}
-|P_{\!\Lambda}^{\rm t}| (m_\varphi^2 M_{1b}+|\transv{p}{}|^2\cos^2\theta M_{3b})
\over
2W_q (-m_\Sigma+P_{\!\Lambda}^\ell+W_q+E_0^\Sigma +m_\varphi)
\big[(m_\Sigma-W_q-P_{\!\Lambda}^\ell)^2-W_p^2\big]   }
\nonumber\\[10pt]&\makebox[0pt]{}&\quad
+\,N_c\,{(m_\Sigma-m_\Lambda\Omega)m_\varphi|\transv{p}{}|\cos\theta M_{2c}
-|P_{\!\Lambda}^{\rm t}| (m_\varphi^2 M_{1c}+|\transv{p}{}|^2\cos^2\theta M_{3c})
\over
\big[(-m_\varphi-E_0^\Sigma+m_\Sigma-P_\Lambda^\ell)^2-W_q^2\big]
\big[(m_\varphi+E_0^\Sigma)^2-W_p^2\big]    } \Bigg\}\,,
\label{baryon-decay-integration-D}
\ea
where $N_a\equiv N(p')|_{\longit{p}=W_p}$\,,
$N_b\equiv N(p')|_{\longit{p}=-m_\Sigma+W_q+P_{\!\Lambda}^\ell}$\,,
$N_c\equiv N(p')|_{\longit{p}=-m_\varphi-E_0^\Sigma}$\,,
$M_{n a}\equiv M_n(p)|_{\longit{p}=W_p}$\,,
$M_{n b}\equiv M_n(p)|_{\longit{p}=-m_\Sigma+W_q+P_{\!\Lambda}^\ell}$\,,
$M_{n c}\equiv M_n(p)|_{\longit{p}=-m_\varphi-E_0^\Sigma}$\,,
$(n=1,2,3)$\,.
Furthermore, $|P_{\!\Lambda}^{\rm t}|\equiv m_\Lambda\sqrt{\Omega^2-1}$ and
$P_{\!\Lambda}^\ell\equiv m_\Lambda\Omega$ are constants. Note that the
energy factors $\sqrt{2E_\Lambda}$ and $\sqrt{2E_\Sigma}$ will be absorbed
into the corresponding BS wave functions. This is convenient since they
appear in the normalization conditions (\ref{normalization-lambda-bsw})
and (\ref{normalization-sigma-bsw}) given in Appendix
\ref{appendix-baryon-bsw-normalization} and make the normalization conditions
be Lorentz invariant.

In the rest frame of $\Sigma_Q^{(*)}$\,, the differential decay width of
$\Sigma_Q^{(*)}\to\Lambda_Q+\pi$ is
\ba
d\Gamma = {1\over 32\pi^2}\,{|{\bf P}_{\!\Lambda}^{}|\over m_\Sigma^2}
|{\cal M}|^2 d\Omega\,,
\ea
where $d\Omega$ is the solid angle of the particle in the final state, and
${\bf P}_{\!\Lambda}^{}$ is the three-momentum of the baryon $\Lambda_Q$\,.
The Lorentz-invariant amplitude in our case can be written as
\ba
{\cal M}= -3 G_{\pi\phi\varphi}^{}\,{\cal D}(q^2)\,\,
\ovl u_\Lambda(v') B^{(m)}_\rho(v)P_{\!\Lambda}^{{\rm t},\rho}\,.
\ea
We will calculate the unpolarized decay width averaging over the spins of
the initial states and summing over the spins of the final states.
For $\Sigma_Q$\,, we have
\ba
{1\over 2}\sum_{s',s}
\big|\ovl u_\Lambda(v',s') B^{(1)}_\rho(v,s)P_{\!\Lambda}^{{\rm t},\rho}\big|^2
={m_\Lambda^2\over 6}\,(\Omega-1)(\Omega+1)^2\,,
\ea
while for $\Sigma_Q^*$\,, we have
\ba
{1\over 4}\sum_{s',s}
\big|\ovl u_\Lambda(v',s') B^{(2)}_\rho(v,s)P_{\!\Lambda}^{{\rm t},\rho}\big|^2
={m_\Lambda^2\over 6}\,(\Omega-1)(\Omega+1)^2\,.
\ea
While deriving the above two equations we have used the following formula for
the spin sum of the Dirac spinor:
\ba
\sum_{s=1,2}u_B(k_B,s)\ovl u_B(k_B,s)  %={{\not\! k_B}+m_B\over 2 m_B}
={{\not\! v_B}+1\over 2}\,,\qquad B=\Lambda_Q \hbox{ or }\Sigma_Q\,,
\ea
with $v_{\Lambda_Q}^{}=v'$, $v_{\Sigma_Q}^{}=v$\,, and the formula for the spin
sum of the Rarita-Schwinger spinor \cite{lurie-book,falk-slac-pub-5689},
\ba
\sum_{s=1}^4 u^\mu(k,s)\ovl u^{\,\nu}(k,s)
%&=&{{\not\! k}+m\over 2 m}\,
%\bigg\{ - \eta^{\mu\nu} + {1\over 3}\gamma^\mu\gamma^\nu
%+ {2\over 3}{k^\mu k^\nu\over k^2}
%+ {{\not\! k}\over 3k^2}(\gamma^\mu k^\nu-\gamma^\nu k^\mu) \bigg\}
%\nonumber\\[5pt]&=&
={{\not\! v}+1\over 2}\,
\bigg\{ - g^{\mu\nu} + {1\over 3}\gamma^\mu\gamma^\nu
+ {2\over 3}v^\mu v^\nu
+ {1\over 3}(\gamma^\mu v^\nu-\gamma^\nu v^\mu) \bigg\}\,,
\ea
where the fact $({\not\! v}+1){\not\! v}={\not\! v}+1$ has been used in the
final step. The total unpolarized decay width in the final form then reads
\ba
\Gamma(\Sigma_Q^{(*)}\to\Lambda_Q+\pi)
={3\over 16\pi}\,|G_{\pi\phi\varphi}|^2\,|{\cal D}(q^2)|^2\,
{|{\bf P}_{\!\Lambda}^{}|\, m_\Lambda^2 \over m_{\Sigma^{(*)}}^2}\,
(\Omega-1)(\Omega+1)^2\,.
\ea

\section{Numerical analysis}
\label{sect: numerical-results}

We will first discuss the parameters appearing in this paper. These
parameters include $m_\phi$, $m_\varphi$, $\kappa_B^{}$, $m_Q$, and $E_0$\,.
The study of mesons in the BS equation approach \cite{meson-in-BS}
shows that the values of the masses of the heavy quarks $m_c=1.58\hbox{\,GeV}$
and $m_b=5.02\hbox{\,GeV}$ lead to predictions in good agreement
with experiment.

Now, we discuss the ranges of the masses of diquarks $\phi$ and
$\varphi$. In contrast with the colorless hadronic states, diquarks (and
other color states) are not free. Therefore, their (effective) masses can not
be measured in experiments. In our discussion, we will treat these masses as
parameters varying in reasonable ranges which satisfy various constraints
from physical considerations.
From the analysis of the spectrum of the heavy baryons,
Ref. \cite{jaffe-diquark-report} shows that
the mass of the scalar diquark is related to that of the axial-vector diquark:
$m_\varphi-m_\phi\approx 0.210\hbox{\,GeV}$
       \footnote{
         $m_\varphi-m_\phi=0.211\hbox{\,GeV}$ for the $c$-baryons and
         $m_\varphi-m_\phi=0.208\hbox{\,GeV}$ for the $b$-baryons
         (the masses of $b$-baryons are taken from Ref. \cite{sigma_b_exp}).
       }.
In our calculation, the mass of $\phi$ is taken to be in the
range $m_\phi\in(0.70,0.80)\hbox{\,GeV}$ and the corresponding range
of the mass of $\varphi$ is $m_\varphi\in(0.91,1.01)\hbox{\,GeV}$.

Now consider $\kappa_B^{}$\,. It is argued in Refs. \cite{guo-1,guo-2}
that the parameter $\kappa_B^{}$ in the effective potential
(See Eq. (\ref{potential-quark-diquark})) can be ranged approximately from
0.02 GeV$^3$ to 0.1 GeV$^3$\,. Furthermore, by studying the average momentum
of $b$-quark in $\Lambda_b$ and comparing with the value of this quantity
derived from the experimental value of the average momentum of the $b$-quark
in the $B$ meson with the aid of HQET, the authors in Ref. \cite{gw-07051379}
show that $\kappa_B^{}$ can be constrained to a narrower range: ``When $m_\phi$
are 0.7 GeV and 0.8 GeV, $\kappa_B^{}$ are roughly in the ranges
$(0.02\hbox{\,-\,}0.06)\hbox{\,GeV}^3$ and
$(0.02\hbox{\,-\,}0.04)\hbox{\,GeV}^3$\,, respectively.'' Following this,
in this paper, we will calculate the decay widths in the range
$\kappa_B^{}\in(0.02\hbox{\,-\,}0.06)\hbox{\,GeV}^3$\,.

Now we will determine $m_\varphi+E_0$. In the heavy quark limit, the baryons
$\Sigma_Q$ and $\Sigma_Q^*$ should be degenerate and the dynamics inside them
are the same. In the heavy quark limit, we can write out the masses of the
baryons as
\ba
m_{\Sigma_Q}^{}=m_Q+m_\varphi+E_0+{\cal O}\left({1\over m_Q}\right)\,,
\ea
where $m_\varphi+E_0$ is the value to the leading order in $1/m_Q$ expansion
and then is universal for all heavy baryons (with one heavy quark). Since
$m_b\gg m_c$, the $1/m_Q$ corrections for $b$-baryons are much smaller than
those for $c$-baryons, hence we shall take the data of $b$-baryons as the input
to calculate this quantity. From the recent results of CDF Collaboration
\cite{sigma_b_exp}, we have
     \footnote{
       We calculate the following quantity by using the spin-averaged
       mass of $\Sigma_b^{(*)}$,
       ${\ovl m}_{\Sigma_b}^{}=(2m_{\Sigma_b}^{}+4 m_{\Sigma_b^*}^{})/6$,
       where $m_{\Sigma_b}^{}=(m_{\Sigma_b^+}^{}+m_{\Sigma_b^-}^{})/2$
       and $m_{\Sigma_b^*}^{}=(m_{\Sigma_b^{+*}}^{}+m_{\Sigma_b^{-*}}^{})/2$.
       Notice that all effects of the isospin symmetry violation are omitted.
     }
\ba
m_\varphi+E_0\approx 0.81\hbox{\,GeV}
\ea
for $\Sigma_Q^{(*)}$ when the $1/m_Q$ corrections are omitted. For
$\Lambda_Q$, we have
\ba
m_\phi+E_0\approx 0.60\hbox{\,GeV}
\ea
when the $1/m_Q$ corrections are omitted.

With the parameters determined above, the decay widths of the processes
$\Sigma_{c,b}^{(*)}\to\Lambda_{c,b}+\pi$ can be obtained. The results are
shown in Table \ref{table: decay width}.
\begin{table}[htb]
\caption{\label{table: decay width} The decay widths
  $\Gamma(\Sigma_{c,b}^{(*)}\to\Lambda_{c,b}+\pi)$ to the leading order in
  $1/m_Q$ expansion. The violation of $SU(2)$ symmetry is not taken
  into account. The unit of $m_\phi$ is GeV, the unit of $\Gamma$ is MeV, the
  unit of $\kappa_B^{}$ is GeV$^3$\,. The mass of $\varphi$ is related to that
  of $\phi$ by $m_\varphi-m_\phi\approx 0.21\hbox{\,GeV}$\,.
}
\center
\begin{tabular}{l|r r r|r r r|r r r}
\hline\hline
$m_\phi$ & & 0.70 & & & 0.75 & & & 0.80 &\\
\hline
$\kappa_B^{}$ & 0.02 & 0.04 & 0.06 & 0.02 & 0.04 & 0.06 & 0.02 & 0.04 & 0.06 \\
\hline
$\Gamma(\Sigma_c)$ & 6.61 & 4.83 & 3.75 & 4.91 & 3.85 & 3.15 & 3.95 & 3.26 & 2.77 \\
$\Gamma(\Sigma_c^*)$ & 18.88 & 14.83 & 12.11 & 17.54 & 14.26 & 12.01 & 15.99 & 13.49 & 11.69\\
\hline
$\Gamma(\Sigma_b)$ & 13.45 & 10.20 & 8.10 & 11.16 & 8.88 & 7.34 & 9.47 & 7.88 & 6.73 \\
$\Gamma(\Sigma_b^*)$ & 17.74 & 13.76 & 11.09 & 15.76 & 12.68 & 10.57 & 13.92 & 11.65 & 10.00\\
\hline\hline
\end{tabular}
\end{table}
From this table, one can see that the theoretical result for
$\Gamma(\Sigma_c^*)$ is consistent with the experimental data \cite{pdg_2006},
$\Gamma^{\rm exp}(\Sigma_c^*)\approx (15\,\hbox{-}\,16)\hbox{\,MeV}$.
However, the theoretical value for $\Gamma(\Sigma_c)$ is bigger than the
experimental data \cite{pdg_2006},
$\Gamma^{\rm exp}(\Sigma_c)\approx 2.2\hbox{\,MeV}$. We
attribute this to the $1/m_Q$ corrections which are not taken into account in
this paper. To look at this more transparently, one can estimate roughly the
$1/m_Q$ corrections as follows. If the corrections to the magnitudes of the BS
wave functions are $\Lambda_{\rm QCD}/m_c= 0.16$ (for
$\Lambda_{\rm QCD}=0.26\hbox{\,GeV}$), the corrections to the final results of
the decay widths can be very large,
$(\Gamma+\delta\Gamma)/\Gamma=1.84\hbox{ to }0.49$ for $c$-baryons. For
$b$-baryons, $\Lambda_{\rm QCD}/m_b= 0.05$, then we have
$(\Gamma+\delta\Gamma)/\Gamma=1.22\hbox{ to }0.81$. Since the corrections for
$b$-baryons are much smaller than those for $c$-baryons we expect that
the predictions for the decay widths of $\Sigma_b^{(*)}$ are far more precise
than those for $\Sigma_c^{(*)}$\,.

\section{Discussions and conclusions}
\label{sect: conclusion}

In this paper, we have first studied the properties of two kinds of diquarks,
the scalar diquark $\phi$ and the axial-vector diquark $\varphi$, in the BS
formalism. We have derived the BS equations for these two kinds of diquarks
and studied all the BS equations under the covariant instantaneous
approximation, which allows one to obtain the BS wave functions in a general
coordinate system directly. With these BS wave functions of the diquarks, we
have calculated the effective coupling constant among the diquarks and the
pion, $G_{\pi\phi\varphi}$. We find that this effective coupling constant is
$|G_{\pi\phi\varphi}|\in (3.35,3.88)$.
With this effective coupling constant, we have calculated the decay widths of
the heavy baryons $\Sigma_Q^{(*)}$ ($Q=c,b$) in the BS formalism in the heavy
quark limit $m_Q\to\infty$.

There are two parameters $m_\phi$ (and $m_\varphi$) and $\kappa_B^{}$ in our model
which can not be determined in this paper. Following the arguments in Ref.
\cite{gw-07051379}, we take
$\kappa_B^{}\in(0.02\hbox{\,-\,}0.06)\hbox{\,GeV}^3$\,.
For the diquark-masses we take
$m_\phi\in[0.70,0.80]\hbox{\,GeV}$ and $m_\varphi\in[0.91,1.01]\hbox{\,GeV}$.
With these ranges of parameters, we give the predictions for the decay widths
of $\Sigma_Q^{(*)}\to\Lambda_Q+\pi$ (the effects of the isospin violation are not
taken into account):
\begin{alignat}{2}
\label{decay-width-sigma-c}
&\Gamma(\Sigma_{c}^{})\approx (2.77\,\hbox{-}\,6.61)\hbox{\,MeV}\,,&\quad
&\Gamma(\Sigma_{c}^*)\approx (11.69\,\hbox{-}\,18.88)\hbox{\,MeV}\,,
\\
&\Gamma(\Sigma_{b}^{})\approx (6.73\,\hbox{-}\,13.45)\hbox{\,MeV}\,,&\quad
&\Gamma(\Sigma_{b}^*)\approx (10.00\,\hbox{-}\,17.74)\hbox{\,MeV}\,.
\label{decay-width-sigma-b}
\end{alignat}
From Eq. (\ref{decay-width-sigma-c}), we can see that the calculated value
for $\Gamma(\Sigma_c^*)$ is consistent with the experimental results
\cite{pdg_2006},
$\Gamma^{\rm exp}(\Sigma_c^*)\approx (15\,\hbox{-}\,16)\hbox{\,MeV}$,
but the calculated value for $\Gamma(\Sigma_c)$ deviates from the experimental
results \cite{pdg_2006}, $\Gamma^{\rm exp}(\Sigma_c)\approx 2.2\hbox{\,MeV}$. We
attribute the deviation to the $1/m_c$ corrections which are not taken into
account in this paper. However, we expect that the predictions for $b$-baryons
are much more precise than those for $c$-baryons since $m_b\gg m_c$.
Furthermore, the above results show that the decay widths of
$\Sigma_b\to\Lambda_b+\pi$ will be much bigger than those of
$\Sigma_c\to\Lambda_c+\pi$.

For comparison, let us quote the results in Ref. \cite{hwang-epjc50-793},
where the decay widths of $\Sigma_b^{(*)}$ were calculated in the bag model
(in units of MeV):
\begin{alignat}{3}
&\Gamma(\Sigma_b^+)=4.35\,, &\qquad &\Gamma(\Sigma_b^0)=5.65\,, &\qquad
&\Gamma(\Sigma_b^-)=5.77\,, \\
&\Gamma(\Sigma_b^{*+})=8.50\,,&\qquad &\Gamma(\Sigma_b^{*0})=10.20\,, &\qquad
&\Gamma(\Sigma_b^{*-})=10.44\,,
\end{alignat}
which are comparable with our results in Eq. (\ref{decay-width-sigma-b}).

In this paper, we omit all the $1/m_Q$ corrections in the calculations. If the
$1/m_Q$ corrections are taken into account, we can take the value of
$\Gamma(\Sigma_c)$, which has been measured precisely in experiments, as the
input to constrain the parameters $\kappa_B^{}$, $m_\phi$, and $m_\varphi$ and
make more precise predictions for $\Gamma(\Sigma_c^*)$ and
$\Gamma(\Sigma_b^{(*)})$.  The full $1/m_Q$ corrections include the $1/m_Q$
corrections to the kernel, to the heavy quark propagators, and to the BS wave
functions. The $1/m_Q$ corrections for $\Lambda_Q$ have been discussed in
Ref. \cite{guo-3}. Since the study of these $1/m_Q$ corrections is very
complicated, this is beyond the scope of the present paper and will be
discussed elsewhere.

Finally, the effect of the isospin violation is not taken into
account to make the presentation of the calculation more transparent. After
taking into account the effect of the isospin symmetry violation one can 
obtain more information on the properties of heavy baryons (if including 
the strange quark $s$, one can also give predictions for $\Omega_Q$ and 
$\Xi_Q$). This will be the subjects in the future work.

\bigskip
\bigskip

\noindent
{\bf Acknowledgments.}
One of us (XHW) is grateful to Dr. Wei Zhang for help on Fortran
programing. This work was supported in part by National Natural Science
Foundation of China (Project Number 10675022), the Key Project of Chinese
Ministry of Education (Project Number 106024) and the Special Grants from
Beijing Normal University.

\appendix

\renewcommand{\theequation}{\thesection\arabic{equation}}

\setcounter{section}{0}
\setcounter{equation}{0}

\section{Some definitions used in the previous sections}
\label{sect-B}

In this appendix, we give definitions of functions used in Subsections
\ref{subsect: bse for scalar diquark} and
\ref{subsect: bse for axial-vector diquark}. The sixteen `matrices' $B_{i j}$
in Subsection \ref{subsect: bse for axial-vector diquark} can be
written as
     \footnote{
       In (and only in) this appendix, $p$ and $p'$ are in fact 
       $\transv{p}{}$ and $\transv{p'}{}$, respectively, which appear in Sect.
       \ref{sect: bs-eq-of-qq}. The change of the notation is to make the 
       expressions more transparent.
     }
\ba
B_{33}&=& -\, 2 \omega^2|p|^2|p'|^2 L_0 + 4 |p|^2|p'|^2 L_1
+ 2(2m^2+\omega^2)L_2
\nonumber\\&&\quad
-\, 4\omega^2|p|^2|p'|^2 G_0 - 16|p|^2|p'|^2 G_1 + 4(2m^2+\omega^2) G_2 \,,
\\[5pt]
B_{34}&=& m m_\varphi(|p|^2|p'|^2 L_0 - 3L_2)\,,
\\[5pt]
B_{35}&=& -\,2m_\varphi^2 |p|^2 (L_1-2G_1)\,,
\\[5pt]
B_{37}&=& -\,4m_\varphi^2|p|^2 (-|p|^2 L_0 +L_1
- 2|p|^2 G_0 - 4 G_1 )   \,,
\ea
%%%%%%%%%%%%%%%%%%%%%%%%%%%%%%%%%%%%%%%%%%%%%%%%%\\[5pt]
\ba
B_{43}&=& -\,{m_\varphi\over m}\Big[ -m^2|p|^2|p'|^2 L_0+2|p|^2|p'|^2 L_1+3m^2L_2
\nonumber\\&&\quad
-2m^2|p|^2|p'|^2 G_0-8|p|^2|p'|^2 G_1+6m^2G_2 \Big]\,,
\\[5pt]
B_{44}&=&-\,2\Big[ m^2|p|^2|p'|^2L_0-(m^2+2\omega^2)L_2\Big]\,,
\\[5pt]
B_{45}&=& 4m m_\varphi|p|^2\Big[-|p|^2L_0+L_1-2G_1 \Big]\,,
\\[5pt]
B_{47}&=& {2m_\varphi^3|p|^2\over m}(L_1-4G_1)    \,,
\ea
%%%%%%%%%%%%%%%%%%%%%%%%%%%%%%%%%%%%%%%%%%%%%%%%%
\ba
B_{53}&=& |p|^2\Big[ |p|^2|p'|^2 L_0 - L_2 + 2|p|^2|p'|^2 G_0 - 2G_2 \Big]\,,
\\[5pt]
B_{54}&=& {2m|p|^2\over m_\varphi} \Big[ -|p|^2|p'|^2 L_0 + L_2 \Big]\,,
\\[5pt]
B_{55}&=& 4|p|^4\Big[ m^2L_0 + L_1 -2G_1\Big]\,,
\\[5pt]
B_{57}&=& -\,2m_\varphi^2|p|^4(L_0+2G_0)\,,
\ea
%%%%%%%%%%%%%%%%%%%%%%%%%%%%%%%%%%%%%%%%%%%%%%%%%\\[5pt]
\ba
B_{73}&=&-\,{2\omega^2|p|^2\over m_\varphi^2}
\Big[ |p|^2|p'|^2 L_0 - L_2 + 2|p|^2|p'|^2 G_0 - 2G_2 \Big]\,,
\\[5pt]
B_{74}&=& {m|p|^2\over m_\varphi}\Big[ |p|^2|p'|^2 L_0 - L_2 \Big]\,,
\\[5pt]
B_{75}&=& -\,2 |p|^4\Big[ m^2L_0 + L_1 -2G_1\Big]\,,
\\[5pt]
B_{77}&=& 4\omega^2|p|^4(L_0+2G_0)\,,
\ea
and the counter terms are
\ba
&&C_{33}=4|p|^4(m^2-|p|^2) L_0\,,\quad C_{34}=-2m m_\varphi|p|^4 L_0\,,
\nonumber\\[5pt]&&
C_{35}=2m_\varphi^2|p|^4 L_0\,, \quad C_{37}=8m_\varphi^2|p|^4 L_0\,,
\\[5pt]&&
C_{43}=-{2m_\varphi |p|^4\over m}(m^2-|p|^2) L_0\,,\quad
C_{44}= 4\omega^2|p|^4 L_0\,,
\nonumber\\[5pt]&&
C_{45}=-8m m_\varphi|p|^4 L_0\,,\quad C_{47}=-{2m_\varphi^3|p|^4\over m} L_0\,,
\ea
%%%%%%%%%%%%%%%%%%%%%%%%%%%%%%%%%%%%%%%%%%%%%%%%%%%%%%
\ba
&&C_{53}=C_{54}=0\,,\quad
C_{55}=4|p|^4(m^2-|p|^2) L_0\,,\quad C_{57}=-2m_\varphi^2|p|^4 L_0\,,
\\[5pt]
&&C_{73}=C_{74}=0\,,\quad
C_{75}=-2|p|^4(m^2-|p|^2) L_0\,,\quad C_{77}=4|p|^4\omega^2 L_0\,,
\ea
where the functions $L_i$ and $G_i$ $(i=0,1,2)$ are defined by
\ba
\int{d^3p'\over (2\pi)^3}\,
{4\pi\kappa\over\left[(p-p')^2+\mu^2\right]^2}\,f(|p'|)
&\equiv& \int d|p'| L_0(|p|,|p'|)\,f(|p'|)\,,
\\[5pt]
\int{d^3p'\over (2\pi)^3}\,
{4\pi\kappa (p \cdot p')\over\left[(p-p')^2+\mu^2\right]^2}\,f(|p'|)
&\equiv& \int d|p'| L_1(|p|,|p'|)\,f(|p'|)\,,
\\[5pt]
\int{d^3p'\over (2\pi)^3}\,
{4\pi\kappa (p \cdot p')^2\over\left[(p-p')^2+\mu^2\right]^2}\,f(|p'|)
&\equiv& \int d|p'| L_2(|p|,|p'|)\,f(|p'|)\,,
\\[5pt]
\int{d^3p'\over (2\pi)^3}\, {2g_{\rm s}^2\over 3}\,
{1\over (p-p')^2+\mu^2}f\,(|p'|)
&\equiv& \int d|p'| G_0(|p|,|p'|)\,f(|p'|)\,,
\\[5pt]
\int{d^3p'\over (2\pi)^3}\, {2g_{\rm s}^2\over 3}\,
{p \cdot p'\over (p-p')^2+\mu^2}f\,(|p'|)
&\equiv& \int d|p'| G_1(|p|,|p'|)\,f(|p'|)\,,
\\[5pt]
\int{d^3p'\over (2\pi)^3}\, {2g_{\rm s}^2\over 3}\,
{(p \cdot p')^2\over (p-p')^2+\mu^2}f\,(|p'|)
&\equiv& \int d|p'| G_2(|p|,|p'|)\,f(|p'|)\,,
\ea
with
\ba
L_0 &\makebox[0pt]{=}&{2\kappa\over\pi}\,
{|p'|^2\over (|p|^2+|p'|^2+\mu^2)^2 - 4|p|^2|p'|^2}\,,
\\[5pt]
L_1 &\makebox[0pt]{=}& -\,{\kappa\over 4\pi}\,{|p'|\over |p|}\, \Bigg\{
{4|p||p'|\,(|p|^2+|p'|^2+\mu^2)
\over (|p|^2+|p'|^2+\mu^2)^2 - 4|p|^2|p'|^2}
+\log{(|p|-|p'|)^2+\mu^2 \over (|p|+|p'|)^2+\mu^2} \Bigg\}\,,
\\[5pt]
L_2 &\makebox[0pt]{=}& {\kappa\over 4\pi}{|p'|\over |p|}\, \Bigg\{
4|p||p'|\,{(|p|^2+|p'|^2+\mu^2)^2 - 2|p|^2|p'|^2\over
(|p|^2+|p'|^2+\mu^2)^2 - 4|p|^2|p'|^2}
\nonumber\\&&\qquad\qquad\qquad
+\,(|p|^2+|p'|^2+\mu^2)
\log{(|p|-|p'|)^2+\mu^2 \over (|p|+|p'|)^2+\mu^2} \Bigg\}\,,
\ea
%%%%%%%%%%%%%%%%%%%%%%%%%%%%%%%%%%%%%%%%%%%%%%%%%%%%%%%%%%%%%%%%%%\\[5pt]
\ba
G_0 &\makebox[0pt]{=}& -\,{\alpha_s\over 3\pi}\,{|p'|\over |p|}\,
\log{(|p|-|p'|)^2+\mu^2 \over (|p|+|p'|)^2+\mu^2}\,,
\\[5pt]
G_1 &\makebox[0pt]{=}& {\alpha_s\over 6\pi}\,{|p'|\over |p|}\,
\Bigg\{ 4|p||p'|
+ (|p|^2+|p'|^2+\mu^2)\log{(|p|-|p'|)^2+\mu^2 \over (|p|+|p'|)^2+\mu^2}
\Bigg\}\,,
\\[5pt]
G_2 &\makebox[0pt]{=}&
-\,{\alpha_s\over 12\pi}\,(|p|^2+|p'|^2+\mu^2)\,{|p'|\over |p|}\,
\Bigg\{ 4|p||p'|
+ (|p|^2+|p'|^2+\mu^2)\log{(|p|-|p'|)^2+\mu^2 \over (|p|+|p'|)^2+\mu^2}
\Bigg\}\,,
\nonumber\\
\ea
where $\alpha_s=g_s^2/(4\pi)$\,, %$f(x)$ is an arbitrary smooth function and
$p\cdot p'=-|p||p'|\cos\theta$\,.

\setcounter{equation}{0}
\section{Some results in Refs. \cite{guo-1,guo-2}}
\label{appendix-ref-guo}

In the previous sections, we have used the BS wave functions of the heavy
baryons $\Lambda_Q$ and $\Sigma_Q^{(*)}$ in Refs. \cite{guo-1,guo-2} as the
input to calculate the decay amplitudes. In this appendix, for readers'
convenience, we write out some expressions (taking from Refs.
\cite{guo-1,guo-2}, but in a different notation) used in this paper explicitly.
For more details, we refer the readers to the original references.

Following Ref. \cite{guo-1}, one can write the BS wave function of
$\Lambda_Q$ to the leading order in $1/m_Q$ expansion in terms of the product
of a scalar function $N(p')$ and the spinor of $\Lambda_Q$,
\ba
\chiM{P_{\!\Lambda}^{}}{}(p') = i \Delta_\phi(-p_2') S(p_1')N(p')u_\Lambda(v') \,,
\ea
where the scalar function $N(p')$ is given by the following integration:
\begin{equation}
N(\longit{p'},|\transv{p'}{}|)
=\int \frac{d^3{k}_{\rm t}{}}{(2\pi)^3}(\widetilde{V}_1
+ 2\longit{p'}\, \widetilde{V}_2)\,
\widetilde{\phi}_{P_{\!\Lambda}^{}}({k}_{\rm t}{})\,.
\label{guo-1-12}
\end{equation}
$\widetilde \phi_{P_{\!\Lambda}^{}}$ depends only on the norm of ${k}_{\rm t}{}$
and has been solved numerically in Ref. \cite{guo-1}. The kernel in the above
equation is defined by
\begin{eqnarray}
\widetilde{V}_1&=&
\frac{8\pi\kappa_B^{}}{[(\transv{p}{}'-{k}_{\rm t}{})^2+\mu^2]^2}-(2\pi)^3
\delta^3  (\transv{p}{}'-{k}_{\rm t}{})
\int \frac{d^3 l}{(2\pi)^3}\frac{8\pi\kappa_B^{}}{(l^2+\mu^2)^2}\,,
\nonumber \\[5pt]
\widetilde{V}_2&=&-\frac{16\pi}{3}
\frac{\alpha_{s,B}^{2}\,Q_{0}^{2}}{[(\transv{p}{}'-{k}_{\rm t}{})^2+\mu^2][(\transv{p}{}'-{k}_{\rm t}{})^2+Q_{0}^{2}]}\,,
\label{potential-quark-diquark}
\end{eqnarray}
where $Q_0^2=3.2\hbox{ GeV}^2$\,, $\mu$ is a parameter which is taken to
be small enough so that the numerical result is insensitive to this
parameter.

Following Ref. \cite{guo-2}, one can write the BS wave functions of
$\Sigma_Q^{(*)}$ to the leading order in $1/m_Q$ expansion in terms of the
product of a tensor function $M^{\lambda\mu}(p)$ and the spinor
of $\Sigma_Q^{(*)}$, $B^{(m)}_{\mu}$ ($m=1,2$, corresponding to $\Sigma_Q$ 
and $\Sigma_Q^*$, respectively),
\ba
\chiM{P_\Sigma}{\lambda}(p) =
{-i\over (\longit{p}+E_0^\Sigma +m_\varphi+i\varepsilon)
(\longit{p}^2-W_{p}^2+i\varepsilon)} \,
M^{\lambda\mu}(p) B^{(m)}_{\mu}(v) \,.
\ea
The tensor function is given by
\ba
M^{\lambda\mu}(p)
= g^{\lambda\mu} M_1(\longit{p},|\transv{p}{}|)
+\,\frac{v^\lambda \transv{p}{}^\mu}{m_\varphi} M_2(\longit{p},|\transv{p}{}|)
-\,\frac{\transv{p}{}^\lambda \transv{p}{}^\mu}{m_\varphi^{2}}
M_3(\longit{p},|\transv{p}{}|)\,,
\label{sigma-tensor-function}
\ea
where
\ba
M_1=\int \frac{d^3 {k_{\rm t}}}{(2\pi)^3}
\Bigg\{ \bigg[\widetilde{A} +\,\widetilde{D}
\frac{(\transv{p}{}\cdot {k_{\rm t}})^2-\transv{p}{}^2 k_{\rm t}^2}
{2\transv{p}{}^2}\bigg]
(\widetilde{V}_1+2\longit{p}\widetilde{V}_2)
-\widetilde{C}
\frac{(\transv{p}{}\cdot {k_{\rm t}})^2-\transv{p}{}^2 k_{\rm t}^2}
{2\transv{p}{}^2}\widetilde{V}_2 \Bigg\}\,,
\label{sigma-tensor-function-M1}
\ea
\ba
M_2&=& {1\over m_\varphi}
\int\frac{d^3 {k_{\rm t}}}{(2\pi)^3}
\Bigg\{\bigg[-\widetilde{A}
+\widetilde{D}\frac{(\transv{p}{}\cdot {k_{\rm t}})^2}{\transv{p}{}^2}\bigg]
\Big[\longit{p}\widetilde{V}_1+(\longit{p}^{2}+m_\varphi^{2})\widetilde{V}_2\Big]
\nonumber\\[5pt]&&~
-\,\widetilde{C}\bigg[(\longit{p}^{2}-m_\varphi^{2})\frac{\transv{p}{}\cdot {k_{\rm t}}}{\transv{p}{}^2}
\widetilde{V}_1+\longit{p} \frac{(\transv{p}{}\cdot {k_{\rm t}})^2}{\transv{p}{}^2}\widetilde{V}_2\bigg] \Bigg\} \,,
\label{sigma-tensor-function-M2}
\ea
\ba
M_3 &=& \int\frac{d^3 {k_{\rm t}}}{(2\pi)^3}\Bigg\{\widetilde{A}
(\widetilde{V}_1+\longit{p}\widetilde{V}_2)
\nonumber\\[5pt]&&~
+\,\widetilde{C}\bigg[\longit{p}\frac{\transv{p}{}\cdot {k_{\rm t}}}{\transv{p}{}^2}\widetilde{V}_1
+\frac{m_\varphi^{2}(3(\transv{p}{}\cdot {k_{\rm t}})^2-\transv{p}{}^2{k_{\rm t}^2})+2\transv{p}{}^2(\transv{p}{}\cdot {k_{\rm t}})^2}{2\transv{p}{}^{4}}\widetilde{V}_2\bigg]
 \nonumber\\[5pt]&&~
+\,\widetilde{D}\bigg[-\frac{m_\varphi^{2}(3(\transv{p}{}\cdot {k_{\rm t}})^2-\transv{p}{}^2
{k_{\rm t}^2})+2\transv{p}{}^2(\transv{p}{}\cdot {k_{\rm t}})^2}{2\transv{p}{}^{4}}
(\widetilde{V}_1+2\longit{p}\widetilde{V}_2)
+\longit{p} \frac{(\transv{p}{}\cdot {k_{\rm t}})^2}{\transv{p}{}^2}\widetilde{V}_2\bigg]
\Bigg\}\,.  \nonumber\\[5pt] \label{sigma-tensor-function-M3}
\ea
The three wave functions, $\widetilde A$, $\widetilde C$, and $\widetilde D$
in Eqs. (\ref{sigma-tensor-function-M1})-(\ref{sigma-tensor-function-M3}) are
scalar functions which depend only on the norm of $k_{\rm t}$ and have been
calculated numerically in Ref. \cite{guo-2}\,.

It is worth pointing out that the following conventions have been used here,
$\transv{p}{}^2=|\transv{p}{}|^2$,
$\transv{p}{}\cdot k_{\rm t}=|\transv{p}{}||k_{\rm t}|$\,, etc., which are
different from our conventions used to discuss the BS equations of
diquarks.

\setcounter{equation}{0}
\section{Normalization of the BS wave functions of the heavy baryons}
\label{appendix-baryon-bsw-normalization}

In this appendix, we will give the normalization conditions for the BS wave
functions of $\Lambda_Q$ and $\Sigma_Q^{(*)}$\,. The normalization conditions
of the BS wave functions of $\Lambda_Q$ and $\Sigma_Q^{(*)}$ have been discussed
in Refs. \cite{guo-1,guo-2}. In order to calculate the decay amplitudes (other
than the weak transition amplitude) of baryons in our case, it is necessary
to obtain the BS wave function for each baryon separately. We start from a
normalization equation which is similar to that of the diquark in Eq.
(\ref{diquark-normalization-condition}). Furthermore, in the heavy quark
limit $m_Q\to\infty$, the normalization conditions in this section should take
the same form as those obtained through the normalization of the Isgur-Wise
functions at the zero recoil point, $\xi(\Omega=1)=1$ \cite{guo-1,guo-2}.
This can be checked with explicit derivation.

\bigskip\bigskip\noindent
{\it Normalization of $\chiM{P_\Lambda^{}}{}$}
\bigskip

\noindent
The normalization equation for the BS wave function of the baryon $\Lambda_Q$ is
given by
\ba
i\int {d^4p\, d^4p'\over(2\pi)^8} \ovlchiM{P_\Lambda^{}}{}(p,s)
\left\{ {\partial\over\partial P_\Lambda^0} \,I_{P_\Lambda^{}}(p,p')\right\}
\chiM{P_\Lambda^{}}{}(p',s') = \delta_{s,s'}\,,
\ea
where $s$ and $s'$ are indices of the spin of the baryon and
\ba
I_{P_\Lambda^{}}(p,p')={1\over 3}\,(2\pi)^4\delta^4(p-p')
S_Q^{-1}(p_1) \Delta_\phi^{-1}(p_2)\,.
\ea
Then we have
\ba
{i\over 3}\int {d^4p\over(2\pi)^4} \ovlchiM{P_\Lambda^{}}{}(p,s)
{\partial\over\partial P_\Lambda^0} \,
\Big\{S_Q^{-1}(p_1) \Delta_\phi^{-1}(p_2) \Big\}
\chiM{P_\Lambda^{}}{}(p,s') = \delta_{s,s'}\,.
\ea
After carrying out the integration over $\longit{p}$\,, the normalization
equation becomes (multiplying $\delta_{s,s'}$ on both sides and summing over
the spin indices)
\ba
{E_\Lambda\over 6m_\Lambda} \int {d^3\transv{p}{}\over(2\pi)^3}\,
{(\widetilde\phi_1-2W_p\widetilde\phi_2)
\over W_p(E_0+m_\phi-W_p)^2}\,
\Big\{\widetilde\phi_1
+2\big[(1-2\lambda_1')W_p-2\lambda_2'(E_0+m_\phi)\big]\widetilde\phi_2
\Big\}
= 1\,,
\nonumber
\ea
which can be reduced further to a one-dimensional integral equation,
\ba
{E_\Lambda\over 6m_\Lambda} \int {d |\transv{p}{}|\over 2\pi^2}\,
{|\transv{p}{}|^2(\widetilde\phi_1-2W_p\widetilde\phi_2)
\over W_p(E_0+m_\phi-W_p)^2}\,
\Big\{\widetilde\phi_1
+2\big[(1-2\lambda_1')W_p-2\lambda_2'(E_0+m_\phi)\big]\widetilde\phi_2
\Big\}  =  1\,,
\nonumber\\ \label{normalization-lambda-bsw}
\ea
where $W_p=\sqrt{|\transv{p}{}|^2+m_\phi^2}$\,, $\widetilde\phi_{1,2}$ are
defined by
\ba
\widetilde\phi_1(\transv{p}{})=\int \frac{d^3{k}_{\rm t}{}}{(2\pi)^3} \,
\widetilde{V}_1(\transv{p}{}-\transv{k}{}) \,
\widetilde{\phi}_{P_{\!\Lambda}^{}}^{}({k}_{\rm t}{})\,,\quad
\widetilde\phi_2(\transv{p}{})=\int \frac{d^3{k}_{\rm t}{}}{(2\pi)^3} \,
\widetilde{V}_2(\transv{p}{}-\transv{k}{}) \,
\widetilde{\phi}_{P_{\!\Lambda}^{}}^{}({k}_{\rm t}{})\,.
\ea
As pointed out in the beginning of this section, in the heavy quark limit,
the normalization condition above should reduce to that obtained by the
normalization of the Isgur-Wise function at the zero recoil point. This can be
checked easily. In the heavy quark limit, $m_Q\to\infty$, we have
$\lambda_1'=1$ and $\lambda_2'=0$, then Eq. (\ref{normalization-lambda-bsw})
becomes
\ba
{E_\Lambda\over 3m_\Lambda} \int {d |\transv{p}{}|\over 2\pi^2}\,
{|\transv{p}{}|^2(\widetilde\phi_1-2W_p\widetilde\phi_2)^2
\over 2W_p(E_0+m_\phi-W_p)^2}
 =  1\,.
\ea
One can see that this is the same equation as that given in Ref. \cite{guo-1}
at the zero recoil point $\Omega=1$ (see Eq. (26) in Ref. \cite{guo-1},
notice that $\Omega$ is written as $\omega$ there)
     \footnote{
       In fact there is an extra factor $E_\Lambda/(3m_\Lambda)$ in this equation
       when compared with Eq. (26) of Ref. \cite{guo-1}. The energy factor
       $E_\Lambda/m_\Lambda$ is required to make the BS wave function
       $\widetilde\phi$ be a Lorentz scalar in our convention of one-particle
       states,
       $\langle {\bf p}|{\bf p}'\rangle=(2\pi)^3\delta^3({\bf p}-{\bf p}')$\,,
       which is different from that used in Ref. \cite{guo-1},
       $\langle {\bf p}|{\bf p}'\rangle
       =(2\pi)^3\delta^3({\bf p}-{\bf p}')E_p/m$\,.
       Furthermore, the factor $1/3$, which comes from the summation of the
       color indices, does not appear in Eq. (26) of Ref. \cite{guo-1}.
       However, the ignorance of this color factor does not affect the results
       when one calculate quantities like those in Ref. \cite{guo-1}\,.
     }.

\bigskip\bigskip\noindent
{\it Normalization of $\chiM{P_\Sigma^{}}{\lambda}$}
\bigskip

\noindent
For the baryon $\Sigma_Q^{(*)}$\,, the normalization equation of the BS wave
function is
\ba
i\int {d^4p\, d^4p'\over(2\pi)^8} \ovlchiM{P_\Sigma^{}}{\mu}(p,s)
\left\{ {\partial\over\partial P_\Sigma^0} \,
I_{P_\Sigma^{}}^{}(p,p')_{\mu\nu}\right\}
\chiM{P_\Sigma^{}}{\nu}(p',s') = \delta_{s,s'}\,,
\label{normalization-sigma-1}
\ea
where
\ba
I_{P_\Sigma^{}}^{}(p,p')_{\mu\nu}={1\over 3}\,(2\pi)^4\delta^4(p-p')
S_Q^{-1}(p_1) \Delta_\varphi^{-1}(p_2)_{\mu\nu}\,.
\ea
Multiplying $\delta_{ss'}$ on both sides of Eq.\,(\ref{normalization-sigma-1})
and summing over $s$ and $s'$\,, the normalization equation becomes (using the
relations given in Appendix \ref{appendix-ref-guo})
\ba
1&\makebox[0pt]{=}&\int{d^3\transv{p}{}\over(2\pi)^3}\,
{\lambda_2(v\cdot\eta)\over 18 m_\varphi^4 W_p^3(W_p-E_0-m_\varphi)^2}
\nonumber\\&&{\hskip -1.5cm}\times\,
\Bigg\{ {\lambda_1\over\lambda_2}\,
W_p^2\Big[-3m_\varphi^4M_1M_1'+m_\varphi^2|\transv{p}{}|^2(M_1M_3'+M_2M_2'-M_3M_1')
+ |\transv{p}{}|^4M_3M_3' \Big]\bigg|_{\longit{p}=-W_p}
\nonumber\\&&{\hskip -1cm} +\,
 3 m_\varphi^4 W_p^2(M_{10}-W_p M_{11})\big[M_{10}+(W_p-2\Delta_B)M_{11}\big]
\nonumber\\[5pt]&&{\hskip -1cm}+\,
m_\varphi^2|\transv{p}{}|^2\Big[
W_p^2(2M_{31}+m_\varphi M_{22})\big[W_p(2\Delta_B-W_p)M_{11}-\Delta_B M_{10}\big]
\nonumber\\[5pt]&&{\hskip -1cm}\quad +\,
W_p^2(M_{10}-\Delta_B M_{11})(2M_{30}+m_\varphi M_{21})
+m_\varphi M_{20}\big[(\Delta_B-2W_p)M_{10}+W_p^2 M_{11})\big]  \Big]
\nonumber\\[5pt]&&{\hskip -1cm}+\,
|\transv{p}{}|^4 \Big[
W_p^2(M_{31}+m_\varphi M_{22})\big[W_p(2\Delta_B-W_p)M_{31}-\Delta_B M_{30}\big]
\nonumber\\&&{\hskip -1cm}\quad +\,
W_p^2(M_{30}-\Delta_B M_{31})(M_{30}+m_\varphi M_{21})
+m_\varphi M_{20}\big[(\Delta_B-2W_p)M_{30}+W_p^2 M_{31}
\big]\Big]\Bigg\}\,,\nonumber\\
\label{normalization-sigma-bsw}
\ea
where $\Delta_B=E_0+m_\varphi$\,, $W_p=\sqrt{|\transv{p}{}|^2+m_\varphi^2}$\,,
$v\cdot\eta=E_\Sigma/m_\Sigma$, $M_n$ ($n$=1,2,3) have been defined in
Eqs.\,(\ref{sigma-tensor-function-M1})-(\ref{sigma-tensor-function-M3}),
and $M_{nm} (nm=10,11,20,21,22,30,31)$ are defined by the expansion of $M_n$'s ,
\ba
M_1=M_{10}+\longit{p}M_{11}\,,\quad
M_2=M_{20}+\longit{p}M_{21}+\longit{p}^2M_{22}\,,\quad
M_3=M_{30}+\longit{p}M_{31}\,.
\ea
$M_n' (n=1,2,3)$ in Eq. (\ref{normalization-sigma-bsw}) are given by 
$M_1'=M_1$, $M_2'=M_{20}$ and 
$M_3'=-(M_{30}+m_\varphi M_{21})-\longit{p}(M_{31}+m_\varphi M_{22})$.
In the heavy quark limit, Eq. (\ref{normalization-sigma-bsw}) reduces to
\ba
1&\makebox[0pt]{=}&\int{d|\transv{p}{}|\over 2\pi^2}\,
{|\transv{p}{}|^2\,(v\cdot\eta)\over 18 m_\varphi^4 W_p(W_p-E_0-m_\varphi)^2}
\nonumber\\&&\times\,
\Big[-3m_\varphi^4M_1M_1'+m_\varphi^2|\transv{p}{}|^2(M_1M_3'+M_2M_2'-M_3M_1')
+ |\transv{p}{}|^4M_3M_3' \Big]\bigg|_{\longit{p}=-W_p}\,,
\nonumber
\ea
which gives the same result as that obtained through the normalization
of Isgur-Wise function at the zero recoil point \cite{guo-2}
(the discussion about the extra factor appearing in this equation is the same
as that given in the previous footnote).

\end{document}